\documentclass[journal,10pt]{IEEEtran}
\usepackage[cmex10]{amsmath}
\usepackage{graphicx}
\usepackage{algorithm}
\usepackage{algorithmic}
\usepackage{stfloats}
\usepackage{epstopdf}
\usepackage{multirow}
\usepackage{color}
\usepackage{balance}
\usepackage{amsmath, amssymb}
\usepackage{amsfonts}
\usepackage{caption}
\usepackage{subfigure}
\usepackage{hyperref}
\usepackage{comment}
\usepackage{mathtools}
\usepackage{amsthm}  

\usepackage{amsthm}  
\newtheorem{lemma}{Lemma}
\usepackage{comment}
\usepackage{url}
\usepackage{bm}
\ifCLASSINFOpdf
\else
\fi
\begin{document}
\title{Joint Beamforming and Position Optimization for
 FIRES-NOMA Assisted Wireless Communication Systems}
\author{Yu~Liu,~\IEEEmembership{Student Member,~IEEE,} Qu~Luo,~\IEEEmembership{Member,~IEEE,}~Gaojie~Chen,~\IEEEmembership{Senior Member,~IEEE,}
\\Pei~Xiao,~\IEEEmembership{Senior Member,~IEEE,}
Ahmed~Elzanaty,~\IEEEmembership{Senior Member,~IEEE,}  Mohsen~Khalily,~\IEEEmembership{Senior Member,~IEEE}\\
and Rahim~Tafazolli,~\IEEEmembership{Fellow,~IEEE}

\thanks{\noindent Yu Liu, Qu Luo, Pei Xiao, Ahmed Elzanaty, Mohsen Khalily, and Rahim Tafazolli are with the Institute for Communication Systems (ICS), 5GIC \& 6GIC, University of Surrey, Guildford, Surrey GU2 7XH, U.K. (e-mail: \{y.u.liu, q.u.luo, p.xiao, a.elzanaty, m.khalily, r.tafazolli\}@surrey.ac.uk). 

Gaojie Chen is with the School of Flexible Electronics (SoFE), Sun Yat-sen University, Sun Yat-sen University, Shenzhen, Guangdong 518107, China (e-mail: gaojie.chen@ieee.org).    
(\textit{Corresponding author: Gaojie Chen}). 
}

}

\maketitle

\begin{abstract} 
To address the limitations of conventional reconfigurable intelligent surfaces (RIS) in spatial control capability, this paper  proposes a fluid integrated reflecting and emitting surface (FIRES) assisted non-orthogonal multiple access (NOMA) multi-user communication system. In this system, each FIRES element can continuously and flexibly adjust its position in response to environmental variations, enabling simultaneous service to users in both transmission and reflection zones. This significantly enhances the system's spatial degrees of freedom (DoF) and service adaptability.
To maximize the system’s total sum rate, we formulate a non-convex optimization problem that jointly optimizes the base station beamforming, the transmission/reflection coefficients of the FIRES, and the element positions. An alternating optimization (AO) algorithm is developed, incorporating successive convex approximation (SCA), semi-definite relaxation (SDR), and majorization-minimization (MM) techniques. In particular, to address the complex channel coupling introduced by the coexistence of direct and FIRES paths, the MM framework is employed in the element position optimization subproblem, enabling an efficient iterative solution strategy.
Simulation results validate that the proposed system achieves up to a 27\% increase in total sum rate compared to conventional STAR-RIS systems and requires approximately 50\% fewer RIS elements to attain the same performance, highlighting its effectiveness for cost-efficient large-scale deployment.
\end{abstract}


\IEEEpeerreviewmaketitle

\section{Introduction}
In the upcoming sixth-generation (6G) networks, the ever-growing demand for massive connectivity under limited spectrum resources calls for higher spectral efficiency and more efficient user access \cite{9509294,10147045}. Reconfigurable intelligent surface (RIS) has emerged as a promising and innovative technology, owing to its ability to reconfigure wireless propagation environments through low-cost passive elements, thereby enhancing coverage, energy efficiency, and interference management \cite{10596064}.
On the other hand, as one of the key technologies proposed in fifth-generation (5G) networks, non-orthogonal multiple access (NOMA) has been extensively investigated for its potential to improve spectrum utilization and achieve massive connectivity.  In particular, in power-domain (PD) NOMA, multiple users share the same spectrum resources and use successive interference cancellation (SIC) to enable multi-user detection \cite{9693417}.  The integration of NOMA with advanced wireless technologies has demonstrated great promise in dense multi-user scenarios \cite{10504837}.

Although conventional RIS enhances communication performance, its reflection-only capability limits full-space coverage and user support. To overcome this, simultaneously transmitting and reflecting reconfigurable intelligent surface (STAR-RIS) has been introduced, enabling each element to simultaneously or selectively transmit and reflect signals \cite{9570143}, thus extending RIS functionality to both sides and enhancing deployment flexibility \cite{10133841}. On this basis, integrating STAR-RIS with NOMA has gained attention due to its potential for improved spectral efficiency. Recent studies have explored STAR-RIS-NOMA in diverse scenarios, including multi-user communication \cite{9863732}, covert transmission \cite{10380743}, ultra-reliable and low-latency communications (URLLC) \cite{10440372}, integrated sensing and communications (ISAC) \cite{10472878}, and full-duplex systems \cite{10820060},  demonstrating its adaptability to complex communication demands.

Despite the great potential of RIS, including STAR-RIS, in reshaping wireless propagation, their structural rigidity remains a key limitation. Most existing designs adopt fixed, uniformly spaced elements, restricting spatial degrees of freedom (DoF) and causing performance bottlenecks in beamforming accuracy, adaptability, and interference suppression \cite{11023100}. Increasing the number of elements can improve array gain, but also leads to excessive channel estimation overhead and circuit power consumption \cite{10480438}. Therefore, achieving higher spatial control with fewer RIS elements has become an important research direction.

In recent years, the fluid antenna system (FAS) has gained increasing attention as a flexible architecture that enables dynamic repositioning of antenna elements within a predefined region \cite{10753482}. This additional spatial degree of freedom enhances channel quality, interference suppression, and user coverage, yielding notable performance gains in various scenarios \cite{10599127,10539238,10949741}.
Inspired by this, researchers have extended the concept to RIS, proposing the fluid-RIS architecture, where each element can move flexibly within a predefined area. By jointly optimizing element positions and phase shifts, fluid-RIS introduces an extra spatial degree of freedom and further improves system performance.

{In existing studies, the authors of \cite{10430366} proposed a mobile RIS architecture inspired by fluid antennas and developed a unified non-uniform discrete phase-shift strategy to enhance system performance under Rician fading channels. In \cite{ye2025joint}, a fluid-RIS-assisted ISAC framework was investigated, where the joint optimization of the symbol detector, transmit beamformer, and the phase shifts and positions of the fluid-RIS elements effectively mitigates sensing mismatch and communication errors, thereby improving both multi-target sensing and multi-user communication performance. Similarly, the authors of \cite{salem2025first} studied single-user SISO and multi-user MISO scenarios by jointly optimizing the positions and phase shifts of fluid-RIS elements to achieve significant throughput gains. Furthermore, a fluid-RIS-assisted physical-layer security framework was examined in \cite{kaveh2025physical}, which analyzed downlink secure transmission under spatial correlation and eavesdropping by deriving analytical bounds on the secrecy outage probability and average secrecy capacity. However, all these works are restricted to reflecting-only RIS architectures.}

{More recently, the concept of a fluid integrated reflecting and emitting surface (FIRES) was introduced in \cite{kkwong2025fires}, combining the reconfigurability of fluid antennas with the dual reflection and transmission capability of STAR-RIS. While this work demonstrates the feasibility of FIRES, it adopts a discrete preset-position structure and focuses on a two-user downlink multicast scenario, where inter-user interference is inherently absent. In addition, the direct links between the base station (BS) and users are neglected, which limits its applicability to more general multi-user systems.}

{Motivated by the advantages of FIRES, this paper investigates a FIRES-assisted NOMA system, referred to as FIRES-NOMA.  Different from the discrete FIRES model in \cite{kkwong2025fires}, we adopt a continuous-position FIRES architecture, where each element can move freely within a predefined region. Recent advances in movable antenna systems and reconfigurable electromagnetic surfaces further indicate that spatially reconfigurable element deployment at the wavelength scale is practically feasible \cite{10286328}, \cite{qu2023}. This shift from discrete position selection to continuous spatial modeling represents a fundamental change in the system formulation and naturally leads to a more general yet challenging design space.  Moreover, unlike multicast-based FIRES designs, we explicitly consider a multi-user scenario with inter-user interference, as well as the coexistence of direct and FIRES-assisted transmission and reflection paths, which further increases the coupling among the involved channels.
Our objective is to maximize the sum rate of the proposed FIRES-NOMA system through the joint optimization of the BS beamforming, the transmission and reflection coefficients, and the positions of the FIRES elements.}

 {The main contributions are summarized as follows:}

\begin{itemize}
    \item {A novel multi-user FIRES architecture is proposed, where each element supports simultaneous transmission and reflection and can move continuously within a predefined area. This enables concurrent service to users in both transmission and reflection zones, enhancing spatial degrees of freedom.}
    
    \item {Introducing NOMA into the FIRES system, a multi-user transmission mechanism coordinating the spatial and power domains is constructed. FIRES actively shapes the equivalent channel difference between users through a reconfigurable propagation environment, while NOMA utilizes this difference in the power domain for successive interference cancellation, thereby achieving efficient multi-user access and interference management.}

    \item We formulate a challenging non-convex joint optimization problem that simultaneously involves the beamforming design at the BS, the transmission/reflection coefficients of the FIRES, and the positions of its elements. To tackle the strong variable coupling, we propose a structured three-layer AO framework, where each subproblem is addressed using appropriate techniques such as semi-definite relaxation (SDR), successive convex approximation (SCA), and majorization-minimization (MM), thus reducing the complexity of the original problem.

    \item To overcome the increased coupling among all involved channels caused by the coexistence of direct and FIRES-assisted transmission/reflection paths, we propose an MM-based position optimization method. This approach constructs a tractable surrogate function and iteratively refines the element positions, enabling efficient resolution of the otherwise intractable non-convex subproblem.

    \item Simulation results demonstrate that the proposed FIRES-NOMA system outperforms existing benchmark schemes in multi-user communication scenarios. Specifically, with the same number of RIS elements, it achieves up to a {27\% increase in total sum rate} compared to conventional STAR-RIS systems. Moreover, to achieve the same total sum rate, the proposed system requires {approximately 50\% fewer RIS elements}, highlighting its potential to reduce deployment cost and system complexity in large-scale STAR-RIS applications.
\end{itemize}

The remainder of this paper is structured as detailed below. The proposed model of the FIRES-NOMA assisted communication system is outlined in Section
II, followed by the
problem formulation of the total sum rate maximization.
Section III illustrates the particulars of the proposed alternating algorithms for optimizing the beamforming of the BS, the reflecting and transmitting coefficients of the FIRES, and the positions of the FIRES elements, respectively. 
Section IV showcases numerical outcomes to assess the efficacy of the proposed algorithm compared with a benchmark scheme. Our conclusion is presented in Section V.

\textit{Notations:} Vectors are denoted by boldface lower-case letters, while matrices are denoted by boldface upper-case letters. The superscripts $(\cdot)^T$ and $(\cdot)^H$ denote the transpose and Hermitian transpose, respectively. The spaces $\mathbb{C}^{m \times n}$ and $\mathbb{R}^{m \times n}$ represent the complex and real matrix spaces of dimensions $m \times n$, respectively. The notations $\text{Tr}(\cdot)$, $\text{Rank}(\cdot)$, $\text{diag}(\cdot)$, and $[\cdot]_{\text{X},j}$ are used to denote the trace, rank, diagonal matrix, and the $(\text{X}, j)$th entry indexed by category $\text{X}$ and index $j$, respectively. The Euclidean norm of a vector is denoted by $\|\cdot\|$, and the absolute value of a scalar is denoted by $|\cdot|$. The expectation operation is represented by $\mathbb{E}\{\cdot\}$, the imaginary part of a complex number by $\Im$, and the real part by $\Re$. Additionally, $\odot$ represents the Hadamard product, and $\text{vec}(\cdot)$ denotes the vectorization operation. The notation $\mathbf{W} \succeq 0$ indicates that $\mathbf{W}$ is positive semidefinite. Lastly, $\mathcal{O}(\cdot)$ is used to denote the big-O computational complexity notation.

\section{System Model and Problem Formulation}
In this section, we first describe the FIRES-NOMA-assisted communication system. Then, we formulate the optimization problem aimed at maximizing the total sum rate.
\subsection{System Model}
\begin{figure}[t]
        \centering
       \includegraphics*[width=100mm]{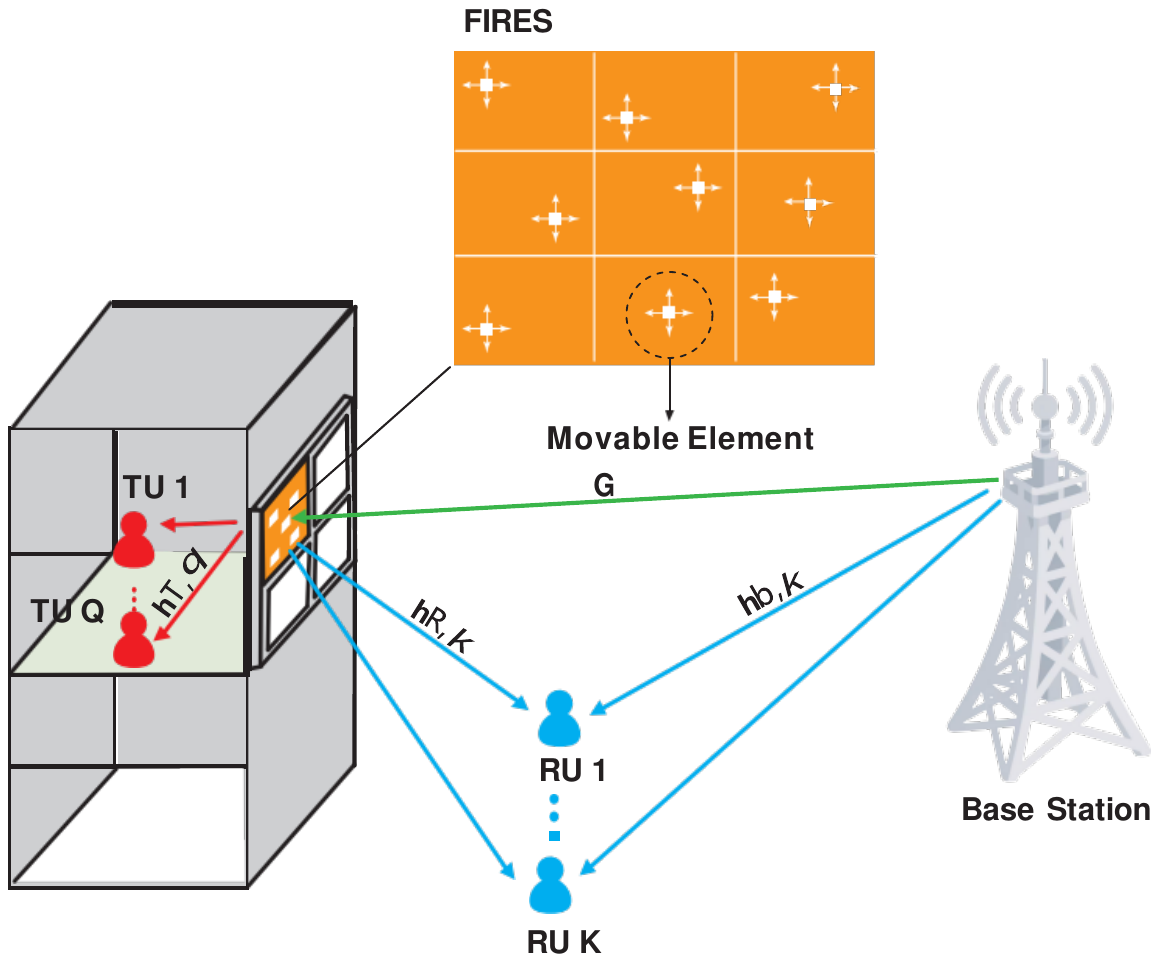}
       \caption{An illustration of FIRES-NOMA assisted multi-user communication system.}
        \label{fig:RIS}
        \vspace{-1em}
\end{figure}

{We consider a FIRES-assisted communication system, where a BS equipped with an \( M \)-antenna uniform linear array (ULA) communicates with multiple single-antenna users via a FIRES comprising \( L \) elements. The ULA is assumed to be a half-wavelength spaced linear array, aligned along the horizontal axis and centered at the origin of the transmit coordinate system.
Each FIRES element is capable of simultaneously reflecting and transmitting incident signals, thereby serving users located in both the reflection and transmission regions. Users in the transmission region are referred to as T users, indexed by \( q \in \{1, 2, \dots, Q\} \), while those in the reflection region are denoted as R users, indexed by \( k \in \{1, 2, \dots, K\} \), as illustrated in Fig.~1.
Moreover, the \( L \) elements of the FIRES are movable and can reposition within a predefined square region \( \mathcal{A} \subset \mathbb{R}^2 \) of size \( A \times A \). The position of the \( l \)th element is denoted by \( \mathbf{p}_l = [p_{x,l}, p_{y,l}]^T \in \mathbb{R}^2 \), and the overall configuration is represented by \( \mathbf{p} = [\mathbf{p}_1, \mathbf{p}_2, \dots, \mathbf{p}_L] \in \mathbb{R}^{2 \times L} \). The considered system focuses on a quasi-static deployment scenario, where user
locations and large-scale channel conditions remain approximately constant
during each FIRES configuration period.
The FIRES operates in the energy splitting (ES) mode \cite{9690478}, where each element divides the incident signal energy between reflection and transmission.}

To characterize the impact of FIRES on the incident signals, the reflecting and transmitting coefficients for the incidence from the BS side are represented as ${\mathbf{\Phi}_{\text{b}}}={\text{diag}}\left\{ {{u_{\text{b},1}}{e^{j{\mu _{\text{b},1}}}},{u_{\text{b},2}}{e^{j{\mu _{\text{b},2}}}}, \ldots ,{u_{\text{b},L}}{e^{j{\mu _{\text{b},L}}}}} \right\} \in {\mathbb{C}^{L \times L}}$ and ${{\mathbf{\Theta}}_{\text{b}}}={\text{diag}}\left\{ {{v_{\text{b},1}}{e^{j{\nu _{\text{b},1}}}},{v_{\text{b},2}}{e^{j{\nu _{\text{b},2}}}}, \ldots ,{v_{\text{b},L}}{e^{j{\nu _{\text{b},L}}}}} \right\}\in{\mathbb{C}^{L \times L}}$, respectively, and $u_{\text{b},l}$ and $v_{\text{b},l}$ represent the reflecting and transmitting amplitudes of the $l$th element, respectively, where $\mu_{\text{b},l}$ and $\nu_{\text{b},l}$ indicate the phase shifts for reflection and transmission at the $l$th element, respectively. Similarly, corresponding matrices for the incidence from user side are represented as ${{\mathbf{\Phi}}_{\text{u}}}={\text{diag}}\left\{ {{u_{\text{u},1}}{e^{j{\mu _{\text{u},1}}}},{u_{\text{u},2}}{e^{j{\mu _{\text{u},2}}}}, \ldots ,{u_{\text{u},L}}{e^{j{\mu _{\text{u},L}}}}} \right\} \in {\mathbb{C}^{L \times L}}$ and ${{\mathbf{\Theta}}_{\text{u}}}={\text{diag}}\left\{ {{v_{\text{u},1}}{e^{j{\nu _{\text{u},1}}}},{v_{\text{u},2}}{e^{j{\nu _{\text{u},2}}}}, \ldots ,{v_{\text{u},L}}{e^{j{\nu _{\text{u},L}}}}} \right\} \in {\mathbb{C}^{L \times L}}$, respectively. 

Based on \cite{Dual-STAR}, the reflection and transmission coefficients are identical on both sides, thus $\mathbf{\Phi}_{\text{b}} = \mathbf{\Phi}_{\text{u}}$ and $\mathbf{\Theta}_{\text{b}} = \mathbf{\Theta}_{\text{u}}$.
Therefore, only $\mathbf{\Phi}_{\text{b}}$ and $\mathbf{\Theta}_{\text{b}}$ are used for simplicity. The FIRES coefficients are computed at the BS, assuming continuous amplitude and phase models, and then used to configure the surface accordingly.
According to \cite{STAR-2}, the constraints can be defined as follows:
\begin{subequations} \small\label{constraint_1}
\begin{equation} \small\label{1a}
u_{\text{b},l}^2 + v_{\text{b},l}^2 \leq 1,
\end{equation}
\begin{equation} \small\label{1b}
0 \leq u_{\text{b},l}, v_{\text{b},l} \leq 1,
\end{equation}
\begin{equation} \small\label{1c}
0 \leq \mu_{\text{b},l}, \nu_{\text{b},l} < 2\pi, \quad \forall l,
\end{equation}
\end{subequations}
\noindent where \eqref{1a} and \eqref{1b} represent the constraints of the total power of transmission and reflection, and constraints of the amplitude, respectively, and \eqref{1c} imposes the range constraints on their corresponding phase shifts.

The downlink signals transmitted from the BS are represented as
\begin{equation} \small
\mathbf{x} = \sum_{k=1}^{K} \mathbf{w}_{\text{R},k} \mathbf{s}_{\text{R},k} + \sum_{q=1}^{Q} \mathbf{w}_{\text{T},q} \mathbf{s}_{\text{T},q},
\end{equation}
where \( \mathbf{w}_{\text{R},k} \in \mathbb{C}^{M \times 1} \) denotes the downlink transmit beamforming vector for the \( k \)th R user in the reflection region, while \( \mathbf{s}_{\text{R},k} \in \mathbb{C} \) represents the corresponding unit-power data symbol, satisfying \( \mathbb{E}\{|\mathbf{s}_{\text{R},k}|^2\} = 1 \). Similarly, \( \mathbf{w}_{\text{T},q} \in \mathbb{C}^{M \times 1} \) denotes the transmit beamforming vector for the \( q \)th T user in the transmission region, and \( \mathbf{s}_{\text{T},q} \in \mathbb{C} \) is the corresponding unit-power data symbol.

For R users in the reflection region, the received signal at the \( k \)th R user\footnote{
For analytical tractability, we assume perfect CSI is available, which facilitates the evaluation of the system’s upper-bound performance. A detailed discussion of channel estimation techniques for FIRES is beyond the scope of this paper; interested readers are referred to \cite{CSI1,CSI2}.
}
 is given by  
\vspace{0.02em}
\begin{equation} \small\label{csignal }
y_{\text{R},k}  = \left( {\mathbf{h}}_{\text{R},k}^H {\mathbf{\Phi}}_{\text{b}}^H {\mathbf{G}} + \mathbf{h}_{\text{b},k} \right) \mathbf{x} + n_{\text{R},k} ,
\end{equation}

\noindent where \( {\mathbf{h}}_{\text{R},k} \in \mathbb{C}^{L \times 1} \) denotes the channel coefficient vector between the FIRES and the \( k \)th R user, \( {\mathbf{G}} \in \mathbb{C}^{L \times M} \) represents the channel coefficient matrix between the BS and the FIRES, and \( \mathbf{h}_{\text{b},k} \in \mathbb{C}^{1 \times M} \) denotes the direct channel coefficient matrix between the BS and the \( k \)th R user. Additionally, \( n_{\text{R},k}  \sim \mathcal{CN}(0, \sigma_{\text{R},k} ^2) \) represents additive white Gaussian noise (AWGN) with zero mean and variance \( \sigma_{\text{R},k} ^2 \).

Then, for the T users in the transmission region, the received signal of the $q$th T user can be expressed as
\vspace{0.02em}
\begin{equation} \small\label{csignal_q}
{{y}_{\text{T},q}} = {\mathbf{h}}_{\text{T},q}^H{{\mathbf{\Theta}}_{\text{b}}^H}{{\mathbf{G}}}{\mathbf{x}} + {n}_{\text{T},q},
\end{equation}

\noindent where ${{\mathbf{h}}_{\text{T},q}} \in {\mathbb{C}^{L \times 1}}$ is the vector of channel coefficients between the FIRES and the $q$th T user, and ${{n}_{\text{T},q}} \sim \mathcal{CN}\left( {0,\sigma _{\text{T},q}^2} \right)$ represents the AWGN with zero mean and variance $\sigma _{\text{T},q}^2$ of the $q$th T user.

The steering vector of the FIRES depends on the positions of the elements \( \mathbf{p} \) as well as the azimuth angle \( \phi \) and the elevation angle \( \psi \). Specifically, the steering vector of the FIRES is expressed as \cite{10243545}:
\begin{equation} \small
\mathbf{a}(\phi, \psi, \mathbf{p}) = \begin{bmatrix} 
e^{j \frac{2\pi}{\lambda} d_1(\phi, \psi, \mathbf{p}_1)}, & 
\cdots, & 
e^{j \frac{2\pi}{\lambda} d_L(\phi, \psi, \mathbf{p}_L)} 
\end{bmatrix}^T.\label{chuchuwen}
\end{equation}

\noindent The term \( d_l(\phi, \psi, \mathbf{p}) \) represents the path difference of signal propagation from direction \( (\phi, \psi) \) between the \( l \)th element and the reference origin, and is given by
\begin{equation} \small
d_l(\phi, \psi, \mathbf{p}_l) = p_{x,l} \sin(\phi) \cos(\psi) + p_{y,l} \sin(\psi).
\end{equation}
\noindent As the FIRES is deployed on a two-dimensional horizontal plane (e.g., \( z = 0 \)), each element's position is limited to the \( x \)- and \( y \)-axes, i.e., \( \mathbf{p}_l = [p_{x,l}, p_{y,l}]^T \). Meanwhile, the steering direction is modeled in three-dimensional space, thus both azimuth (\( \phi \)) and elevation (\( \psi \)) angles are required.

On the other hand, since the BS is equipped with a ULA, the transmit steering vector \( \mathbf{a}_{\text{t}} \) in the direction of \( \phi_{\text{t}} \) is given by  
\begin{equation} \small
\mathbf{a}_{\text{t}} = \begin{bmatrix} 
e^{j \frac{2\pi}{\lambda} d_0}, & e^{j \frac{2\pi}{\lambda} d_1}, & \cdots, & e^{j \frac{2\pi}{\lambda} d_{M-1}}
\end{bmatrix}^T,
\end{equation}  
where \( d_m = \frac{\lambda}{2} m \sin(\phi_{\text{t}}) \), for \( m = 0,1, \dots, M-1 \). 

The channels from the BS to both the FIRES and the users are modeled as {Rician fading channels} with a dominant line-of-sight (LoS) component, which is typical for far-field propagation scenarios where moderate scattering arises due to environmental complexity. Specifically, the channel from the BS to the FIRES is given by
\begin{equation}\small
\mathbf{G}
=
\sqrt{\zeta_{\mathrm{G}}}
\left(
\sqrt{\frac{\kappa}{\kappa+1}}\,
\mathbf{a}_{\mathrm{r}}(\phi_{\mathrm{r}},\psi_{\mathrm{r}},\mathbf{p})
\mathbf{a}_{\mathrm{t}}^{H}(\phi_{\mathrm{t}})
+
\sqrt{\frac{1}{\kappa+1}}\,
\mathbf{G}_{\mathrm{NLoS}}
\right),
\label{7}
\end{equation}
where \(\kappa\) denotes the Rician factor, \(\zeta_{\text{G}}\) represents the large-scale path loss, and \(\mathbf{G}_{\text{NLoS}}\) models the non-line-of-sight (NLoS) component follows a Rayleigh fading model characterized by zero-mean complex Gaussian entries with unit variance.

{Likewise, the direct channel from the BS to the $k$th R user  \(\mathbf{h}_{\text{b},k} \in \mathbb{C}^{1 \times M}\) is expressed as
\begin{equation}\small
\mathbf{h}_{\text{b},k}
= \sqrt{\zeta_{\mathrm{b},k}}
\left(
\sqrt{\frac{\kappa}{\kappa+1}}\, \mathbf{a}_{\mathrm{t}}^H(\phi_{\mathrm{b},k})
+ \sqrt{\frac{1}{\kappa+1}}\, \mathbf{h}_{\mathrm{NLoS},k}
\right),
\label{eq:Hbk_rician}
\end{equation}
where \(\zeta_{\text{b},k}\) and \(\phi_{\text{b},k}\) denote the path loss and the azimuth angle to the \(k\)th user, respectively, and \(\mathbf{h}_{\text{NLoS},k}\) follows a Rayleigh fading model characterized by zero-mean complex Gaussian entries with unit variance.}

The channels from the FIRES to the R and T users are modeled as deterministic LoS links. Specifically, the channel from the FIRES to the \( k \)th R user is given by
\begin{equation} \small
\mathbf{h}_{\text{R},k}^H = \sqrt{\zeta_{\text{R},k}} \, \mathbf{a}_{\text{R},k}^H(\phi_{\text{R},k}, \psi_{\text{R},k}, \mathbf{p}), \quad \forall k = 1, 2, \dots, K, \label{9}
\end{equation}
and similarly, the channel to the \( q \)th T user is expressed as
\begin{equation} \small
\mathbf{h}_{\text{T},q}^H = \sqrt{\zeta_{\text{T},q}} \, \mathbf{a}_{\text{T},q}^H(\phi_{\text{T},q}, \psi_{\text{T},q}, \mathbf{p}), \quad \forall q = 1, 2, \dots, Q.
\end{equation}
Here, \( \zeta_{\text{R},k} \) and \( \zeta_{\text{T},q} \) denote the large-scale path loss components, while \( \mathbf{a}_{\text{R},k}(\cdot) \) and \( \mathbf{a}_{\text{T},q}(\cdot) \) represent the FIRES steering vectors toward the corresponding users, defined in terms of their azimuth and elevation angles.

{According to the NOMA principle, users with stronger effective channel
conditions are capable of employing successive interference cancellation (SIC)
to decode the signals intended for weaker users prior to decoding their own.
In this work, we consider a commonly adopted and representative deployment
scenario, where users in the reflection region generally experience stronger
effective channel gains than those in the transmission region, due to the
availability of a direct BS--user link in addition to the FIRES-assisted path.
This assumption is introduced to specify a representative SIC decoding order
for system-level analysis and does not imply a universally valid channel
ordering.}
Accordingly, assume the order of the channel gains of the users is given by
\begin{equation} \small
||\mathbf{F}_{\text{T},1}||^2 \leq \cdots \leq ||\mathbf{F}_{\text{T},Q}||^2 \leq ||\mathbf{F}_{\text{R},1}||^2  \leq \cdots \leq ||\mathbf{F}_{\text{R},K}||^2 ,
\end{equation} 
where $\mathbf{F}_{\text{R},k}  =  {\mathbf{h}}_{\text{R},k}^H {\mathbf{\Phi}}_{\text{b}}^H {\mathbf{G}} + \mathbf{h}_{\text{b},k} $, and $\mathbf{F}_{\text{T},q} = {\mathbf{h}}_{\text{T},q}^H{{\mathbf{\Theta}}_{\text{b}}^H}{{\mathbf{G}}}$.
Thus, the SIC decoding order at the $q$th T user  can be expressed as
\begin{equation} \small\begin{aligned}
&|\mathbf{F}_{\text{T},q}\mathbf{w}_{\text{R},K}|^2 \leq \cdots \leq |\mathbf{F}_{\text{T},q}\mathbf{w}_{\text{R},1}|^2 \leq |\mathbf{F}_{\text{T},q}\mathbf{w}_{\text{T},Q}|^2\\&  \leq \cdots \leq |\mathbf{F}_{\text{T},q}\mathbf{w}_{\text{T},1}|^2 ,
\label{SIC1}\end{aligned}\end{equation}
and the SIC decoding order at the $k$th R user  is given by
\begin{equation} \small\begin{aligned}
&|\mathbf{F}_{\text{R},k}\mathbf{w}_{\text{R},K}|^2 \leq \cdots \leq |\mathbf{F}_{\text{R},k}\mathbf{w}_{\text{R},1}|^2 \leq |\mathbf{F}_{\text{R},k}\mathbf{w}_{\text{T},Q}|^2\\&  \leq \cdots \leq |\mathbf{F}_{\text{R},k}\mathbf{w}_{\text{T},1}|^2.
\label{SIC2}\end{aligned}\end{equation}
Therefore, the signal-to-interference-plus-noise-ratio (SINR) of the $q$th T user  can be represented as 
 \begin{equation} \small
\gamma_{\text{T},q} = \frac{|\mathbf{F}_{\text{T},q}\mathbf{w}_{\text{T},q}|^2}{ {\sum\limits_{j = q+1}^Q |\mathbf{F}_{\text{T},q}\mathbf{w}_{\text{T},j}|^2}+{\sum\limits_{k =1}^K |\mathbf{F}_{\text{T},q}\mathbf{w}_{\text{R},k}|^2}+ {\sigma _{\text{T},q}^2}},
\end{equation}  
where $q=1, 2, \cdots,Q-1$, and
 \begin{equation} 
\gamma_{\text{T},Q} = \frac{|\mathbf{F}_{\text{T},Q}\mathbf{w}_{\text{T},Q}|^2}{ {\sum\limits_{k =1}^K |\mathbf{F}_{\text{T},Q}\mathbf{w}_{\text{R},k}|^2}+ {\sigma _{\text{T},Q}^2}}.
\end{equation} 

Accordingly, the SINR of the $k$th R user  can be represented as 
\begin{equation}
\gamma_{\text{R},k} = \frac{|\mathbf{F}_{\text{R},k} \mathbf{w}_{\text{R},k}|^2}{ {\sum\limits_{j=k+1}^K |\mathbf{F}_{\text{R},k} \mathbf{w}_{\text{R},j}|^2}+ {\sigma _{\text{R},k}^2}},
\end{equation}  
where $k=1, 2, \cdots,K-1$, and
 \begin{equation} \small
\gamma_{\text{R},K} = \frac{|\mathbf{F}_{\text{R},K} \mathbf{w}_{\text{R},K}|^2}{{\sigma _{\text{R},K}^2}}.
\end{equation}  

\subsection{Problem Formulation}
The objective of this paper is to maximize the total sum rate of the proposed system by jointly optimizing the BS transmit beamforming, the FIRES coefficients, and the positions of its elements. The optimization is subject to the BS transmit power constraint, the unit-modulus and power constraints of the FIRES, and the physical limitations on element mobility. The resulting problem is formulated as{\small
\begin{align}\small
&\mathop {\max }\limits_{\boldsymbol{\Omega}}\ \ \sum\limits_{k = 1}^K {\log_2 (1+\gamma_{\text{R},k})} +\sum\limits_{q = 1}^Q {\log_2 (1+\gamma_{\text{T},q})}  \label{P1_OF}\\
{\text{s.t.}}\ \ & \ \   {\text{Tr}}\left(\sum _{k=1}^{K}  {{\mathbf{w}_{\text{R},k}}{\mathbf{w}_{\text{R},k}^H}} +\sum\limits_{q = 1}^Q {\mathbf{w}_{\text{T},q}}{\mathbf{w}_{\text{T},q}^H} \right) \leq P_{\text{max}}, \tag{\ref{P1_OF}{a}}  \label{P1a} \\
& \ \  
\mathbf{p}_l\in \mathcal{A}, \quad \forall l = 1, 2, \dots, L
,  \label{P1b} \tag{\ref{P1_OF}{b}}   \\&\ \
\|\mathbf{p}_l - \mathbf{p}_{l'}\|_2 \geq \Delta D, \forall l \ne l' \in \{1, 2, \dots,L\}, \label{P1c}
\tag{\ref{P1_OF}{c}} \\
& \ \  \gamma_{\text{R},k} \geq \gamma_{\text{min}},  \quad \forall k = 1, 2, \dots, K,\tag{\ref{P1_OF}{d}}  \label{P1d} \\
& \ \  \gamma_{\text{T},q} \geq \gamma_{\text{min}},  \quad \forall q = 1, 2, \dots, Q, \tag{\ref{P1_OF}{e}}  \label{P1e} \\
& \ \  \eqref{SIC1}, \eqref{SIC2}, \tag{\ref{P1_OF}{f}}  \label{P1f} \\ 
& \ \  \eqref{1a}-\eqref{1c}, \tag{\ref{P1_OF}{g}}  \label{P1g}
\end{align}}

 \noindent
where \( \boldsymbol{\Omega} \triangleq \left\{ \{\mathbf{w}_{\text{R},k}\}_{k=1}^{K}, \{\mathbf{w}_{\text{T},q}\}_{q=1}^{Q}, \boldsymbol{\Phi}_{\text{b}}, \boldsymbol{\Theta}_{\text{b}}, \mathbf{p} \right\} \) denotes the set of optimization variables. Constraint~\eqref{P1a} limits the total transmit power at the BS to \( P_{\text{max}} \), while \eqref{P1b} confines each RIS element to a predefined planar region. Constraint~\eqref{P1c} ensures a minimum inter-element spacing of \( \Delta D \). Constraints~\eqref{P1d} and \eqref{P1e} impose minimum SINR requirements for all users. The SIC feasibility is enforced by~\eqref{P1f}, and \eqref{P1g} defines the physical constraints on the FIRES coefficients.

\vspace{0.02em}
\section{Proposed Optimization Algorithm}
In this section, we focus on solving problem~\eqref{P1_OF}, which is a non-convex and NP-hard optimization problem due to the joint optimization of multiple coupled variables, making it difficult to tackle directly. Therefore, we divide it into three sub-problems: optimizing the downlink beamforming, optimizing the FIRES coefficients, and optimizing the positions of FIRES elements.
\subsection{ Transmit Beamforming Optimization}
In this subsection,  we optimize the beamforming vectors $\mathbf{\{{w}}_{\text{R},k} \}_{k=1}^{K}$ and $\{\mathbf{{w}}_{\text{T},q}\}_{q=1}^{Q}$  when the FIRES coefficients $\mathbf{\Phi}_b$, $\mathbf{\Theta}_b$, and the FIRES element positions \( \mathbf{p} \) are fixed. By omitting terms that are not coupled with transmit beamforming and defining  ${\mathbf{W}_{\text{R},k}} =\mathbf{w}_{\text{R},k}\mathbf{w}_{\text{R},k}^H,{\mathbf{W}_{\text{R},j}} =\mathbf{w}_{\text{R},j}\mathbf{w}_{\text{R},j}^H$, ${\mathbf{W}_{\text{T},q}} =\mathbf{w}_{\text{T},q}\mathbf{w}_{\text{T},q}^H $, and ${\mathbf{W}_{\text{T},j}} =\mathbf{w}_{\text{T},j}\mathbf{w}_{\text{T},j}^H $,  we employ the semidefinite relaxation (SDR) \cite{5447068} technique to handle the non-convexity, and subsequently reformulate the sub-problem as
\vspace{0.02em}
\begin{equation} \small
\begin{aligned}
\mathop {\max }\limits_{\boldsymbol{\Omega}_{1}} \quad  \sum_{k = 1}^{K} \log_2 \left(1+\frac{\xi_{\text{R},k}}{\psi_{\text{R},k}} \right) 
 + \sum_{q = 1}^{Q} \log_2 \left(1+\frac{\xi_{\text{T},q}}{\psi_{\text{T},q}} \right) 
\end{aligned}
\label{P1_OF2}
\end{equation}
\vspace{-0.1em}
{\small\begin{align}\small
\text{s.t.} \quad & \text{Tr} \left( \sum_{k=1}^{K} \mathbf{W}_{\text{R},k} + \sum_{q = 1}^Q \mathbf{W}_{\text{T},q} \right) \leq P_{\text{max}}, && \tag{\ref{P1_OF2}a} \label{P1_OF2_a} \\
& \frac{\xi_{\text{R},k}}{\psi_{\text{R},k}} \geq \gamma_{\text{min}}, \quad \forall k = 1, 2, \dots, K,&& \tag{\ref{P1_OF2}b} \\
& \frac{\xi_{\text{T},q}}{\psi_{\text{T},q}} \geq \gamma_{\text{min}}, \quad \forall q = 1, 2, \dots, Q,&& \tag{\ref{P1_OF2}c} \\
& \mathbf{F}_{\text{T},q}\mathbf{W}_{\text{R},K}\mathbf{F}_{\text{T},q}^H\leq \cdots \leq \mathbf{F}_{\text{T},q}\mathbf{W}_{\text{T},1}\mathbf{F}_{\text{T},q}^H ,&& \tag{\ref{P1_OF2}d} \\
& \mathbf{F}_{\text{R},k}\mathbf{W}_{\text{R},K}\mathbf{F}_{\text{R},k}^H\leq \cdots \leq \mathbf{F}_{\text{R},k}\mathbf{W}_{\text{T},1}\mathbf{F}_{\text{R},k}^H, && \tag{\ref{P1_OF2}e} \label{P1_OF2_g} 
\end{align}}
\noindent
where $\boldsymbol{\Omega}_{\text{1}} \triangleq \left\{ \{{\mathbf{W}}_{\text{R},k} \}_{k=1}^{K},\{\mathbf{W}_{\text{T},q}\}_{q=1}^{Q} \right\}$, 
$\xi_{\text{R},k} = \mathbf{F}_{\text{R},k} \mathbf{W}_{\text{R},k}\mathbf{F}_{\text{R},k}^H$, a unified notation is adopted for all \( k \) since the expression remains the same.
$\psi_{\text{R},k} = \sum\nolimits_{j = k+1}^K \mathbf{F}_{\text{R},k} \mathbf{W}_{\text{R},j}\mathbf{F}_{\text{R},k}^H + \sigma_{\text{R},k}^2$, $\forall k = 1, 2, \dots,K-1$, 
$\psi_{\text{R},K} = \sigma_{\text{R},K}^2$, 
$\xi_{\text{T},q} = \mathbf{F}_{\text{T},q}\mathbf{W}_{\text{T},q}\mathbf{F}_{\text{T},q}^H$, and a unified notation is adopted for all \( q \) since the expression remains the same.
$\psi_{\text{T},q} = \sum\nolimits_{j = q+1}^Q \mathbf{F}_{\text{T},q}\mathbf{W}_{\text{T},j}\mathbf{F}_{\text{T},q}^H + \sum\nolimits_{k=1}^K \mathbf{F}_{\text{T},q}\mathbf{W}_{\text{R},k}\mathbf{F}_{\text{T},q}^H + \sigma_{\text{T},q}^2$, 
$\forall q = 1, 2, \dots,Q-1,$ and
$\psi_{\text{T},Q} = \sum\nolimits_{k=1}^K \mathbf{F}_{\text{T},Q}\mathbf{W}_{\text{R},k}\mathbf{F}_{\text{T},Q}^H + \sigma_{\text{T},Q}^2$.
Note that the rank-one constraint has been omitted. However, problem~\eqref{P1_OF2} remains non-convex due to the fractional form of its objective function. To address this, we adopt the inequality provided in (72) of~\cite{budengshi}, which offers a tight concave lower bound for functions of the form \( \ln(1 + \xi/\psi) \). Specifically, the following inequality holds:
\vspace{0.2em}
\begin{equation} \small
\begin{aligned}
\ln \left (1 + \frac{\xi}{\psi} \right )  \geq \ln \left (1 + \frac{\xi^{(n)}}{\psi^{(n)}} \right ) + \frac{ \frac{\xi^{(n)}}{\psi^{(n)}} }{1 + \frac{\xi^{(n)}}{\psi^{(n)}}} \left (2 - \frac{\xi^{(n)}}{\xi} - \frac{\psi}{\psi^{(n)}} \right ), \label{budengshi}
\end{aligned}
\end{equation}
where \( \xi^{(n)} \) and \( \psi^{(n)} \) denote the values of \( \xi \) and \( \psi \) obtained in the previous iteration, respectively.

Applying this inequality yields a concave lower-bound approximation of the objective function:
\vspace{0.2em}
\begin{equation} \small
\begin{aligned}
&\sum _{k=1}^{K}\log_2 \left (1 + \frac{\xi _{\text{R},k}^{(n)}}{\psi _{\text{R},k}^{(n)}} \right ) +\sum _{q=1}^{Q}\log_2 \left (1 + \frac{\xi _{\text{T},q}^{(n)}}{\psi _{\text{T},q}^{(n)}} \right ) \\
&+ \sum _{k=1}^{K}\frac{ \frac{\xi _{\text{R},k}^{(n)}}{\psi _{\text{R},k}^{(n)}} }{\ln 2 \left(1 + \frac{\xi _{\text{R},k}^{(n)}}{\psi _{\text{R},k}^{(n)}}\right)} \left (2 - \frac{\xi _{\text{R},k}^{(n)}}{\xi _{\text{R},k}} - \frac{\psi _{\text{R},k}}{\psi _{\text{R},k}^{(n)}} \right )\\
&+ \sum _{q=1}^{Q}\frac{ \frac{\xi _{\text{T},q}^{(n)}}{\psi _{\text{T},q}^{(n)}} }{\ln 2 \left(1 + \frac{\xi _{\text{T},q}^{(n)}}{\psi _{\text{T},q}^{(n)}}\right)} \left (2 - \frac{\xi _{\text{T},q}^{(n)}}{\xi _{\text{T},q}} - \frac{\psi _{\text{T},q}}{\psi _{\text{T},q}^{(n)}} \right ). \label{17}
\end{aligned}
\end{equation}

Therefore, the problem  \eqref{P1_OF2} can be expressed as {\small
\begin{align}\small
&\mathop {\max }\limits_{\boldsymbol{\Omega}_{\text{1}}} \ \ \eqref{17} \label{P1_OF3}\\
{\text{s.t.}}\ \ 
& \ \ \eqref{P1_OF2_a}-\eqref{P1_OF2_g}, \tag{\ref{P1_OF3}{a}}
\end{align}}

\noindent which is convex, and the corresponding optimal solutions can be obtained by convex optimization tools, such as CVX \cite{grant2009cvx}. Built upon the rank reduction theorem in Theorem 3.2 of \cite{rank1}, the problem \eqref{P1_OF3} has an optimal solution $\{\mathbf{\{{W}}_{\text{R},k}^* \}_{k=1}^{K},\{\mathbf{{W}}_{\text{T},q}^* \}_{q=1}^{Q}\}\}$  satisfying the rank-one constraint. Thus, the optimal $\{{\mathbf{\{{w}}_{\text{R},k}^* }\}_{k=1}^{K},\{\mathbf{w}_{\text{T},q}^*\}_{q=1}^{Q}\}$ can be recovered by eigenvalue decomposition.

\subsection{Optimization of the Coefficients for the FIRES }
In this subsection, we optimize the FIRES coefficients \( \mathbf{\Phi}_{\text{b}} \) and \( \mathbf{\Theta}_{\text{b}} \), with the beamforming vectors \( \{ \mathbf{w}_{\text{R},k} \}_{k=1}^{K} \), \( \{ \mathbf{w}_{\text{T},q} \}_{q=1}^{Q} \), and the element positions \( \mathbf{p} \) fixed. Ignoring terms independent of the FIRES coefficients, the subproblem is formulated as

\vspace{-0.01em}{\small
\begin{align}\small
\mathop {\max }\limits_{\mathbf{\Phi}_{\text{b}}, \mathbf{\Theta}_{\text{b}}}&\ \ \sum\limits_{k = 1}^K {\log_2 (1+\gamma_{\text{R},k})} +\sum\limits_{q = 1}^Q {\log_2 (1+\gamma_{\text{T},q})}  \label{P2_OF}\\
{\text{s.t.}}\ \ & \ \  \eqref{P1d}-\eqref{P1f} .\tag{\ref{P2_OF}{a}}  
\end{align}}

\noindent Since the objective function and the constraints are not convex, new variables $\mathbf{v}_1=[{{v_{b,1}}{e^{j{\nu _{b,1}}}},{v_{b,2}}{e^{j{\nu _{b,2}}}}, \ldots ,{v_{b,L}}{e^{j{\nu _{b,L}}}}}]^T$ and  $\mathbf{v}_2 =[{{u_{b,1}}{e^{j{\mu _{b,1}}}},{u_{b,2}}{e^{j{\mu _{b,2}}}}, \ldots ,{u_{b,L}}{e^{j{\mu _{b,L}}}}}] ^T$ are introduced. Firstly, we define
$
\mathbf{u}_{1,q} = \operatorname{diag}(\mathbf{h}_{\text{T},q}^H)\, {\mathbf{G}}\,\mathbf{w}_{\text{T},q}, 
\mathbf{u}_{2,j} = \operatorname{diag}(\mathbf{h}_{\text{T},q}^H)\, {\mathbf{G}}\,\mathbf{w}_{\text{T},j}, 
\mathbf{u}_{3,k} = \operatorname{diag}(\mathbf{h}_{\text{T},q}^H)\, {\mathbf{G}}\,\mathbf{w}_{\text{R},k}, 
$
It is worth noting that the variation in the subscript of \( \mathbf{u}_{i,\cdot} \) only reflects the change in the beamforming vector \( \mathbf{w} \), while the core channel term \( \operatorname{diag}(\mathbf{h}_{\text{T},q}^H)\, {\mathbf{G}} \) remains unchanged as it is solely determined by the reference user \( q \).
 Based on the properties of matrix operations, the terms 
\( |\mathbf{F}_{\text{T},q} \mathbf{w}_{\text{T},q}|^2 \), 
\( |\mathbf{F}_{\text{T},q} \mathbf{w}_{\text{T},j}|^2 \), and 
\( |\mathbf{F}_{\text{T},q} \mathbf{w}_{\text{R},k}|^2 \) 
can be equivalently expressed as 
\( \mathbf{v}_1^H \mathbf{U}_{1,q} \mathbf{v}_1 \), 
\( \mathbf{v}_1^H \mathbf{U}_{2,j} \mathbf{v}_1 \), and 
\( \mathbf{v}_1^H \mathbf{U}_{3,k} \mathbf{v}_1 \), respectively, 
where \( \mathbf{U}_{i,x} = \mathbf{u}_{i,x} \mathbf{u}_{i,x}^H \) for \( (i,x) \in \{ (1,q), (2,j), (3,k) \} \).

To address the non-convexity of the objective function, we first reformulate the term \( \sum\limits_{q = 1}^Q \log_2 (1+\gamma_{\text{T},q}) \) as
$ \sum_{q = 1}^{Q} \log_2 \left(1+\frac{\beta_{\text{T},q}}{\chi _{\text{T},q}}\right)\label{sumq}$,
where $\beta_{\text{T},q} \!= \!\mathbf{v}_1^H \mathbf{U}_{1,q} \mathbf{v}_1$,\! and a unified notation is adopted for all \( q \)\! .
$\chi_{\text{T},q} \!\!=\! \sum\nolimits_{j = q+1}^{Q} \mathbf{v}_1^H \mathbf{U}_{2,j} \mathbf{v}_1 \!+\! \sum\nolimits_{k=1}^{K} \mathbf{v}_1^H \mathbf{U}_{3,k} \mathbf{v}_1 +\sigma_{\text{T},q}^2$,\!\!\!\!\! \quad
$\forall q = 1, 2, \dots, Q-1$, 
$\chi_{\text{T},Q} = \sum\limits_{k=1}^{K} \mathbf{v}_1^H \mathbf{U}_{3,k} \mathbf{v}_1 + \sigma_{\text{T},Q}^2$.

Note that $\beta_{\text{T},q} $  is convex instead of concave. In fractional programming, a concave numerator is generally easier to handle. Therefore, the SCA method, which handles ${\beta_{\text{T},q}} $ by its first-order Taylor expansion \cite{lipp2016variations}, is applied. Specifically, the following surrogate function is considered:
\begin{equation} \small\begin{aligned}
\beta_{\text{T},q} 
\approx 2\Re{\{{\left({{\mathbf{v}}_\mathrm{{1}}^{(t)}} \right)^{H}}{\mathbf{U}_{1,q}}\}}{{\mathbf{v}}_\mathrm{{1}}}+ {\left({{\mathbf{v}}_\mathrm{{1}}^{(t)}} \right)^{H}}{\mathbf{U}_{1,q}}{{\mathbf{v}}_\mathrm{{1}}^{(t)}},\label{t1}
\end{aligned}\end{equation}
where $\hat{{\mathbf{v}}}_\mathrm{{1}}^{(t)}$ represents a feasible point at the $t$th iteration. Note that after each iteration, $\hat{{\mathbf{v}}}_\mathrm{{1}}^{(t)}$ will be updated according to the optimal solutions obtained.

According to \eqref{budengshi} and \eqref{t1}, the term  \( \sum\limits_{q = 1}^Q \log_2 (1+\gamma_{\text{T},q}) \) can be rewritten as 
\begin{equation}  
\small
\begin{aligned}
\sum _{q=1}^{Q}\log_2 \left ({{1 \! +\! \frac {\beta_{\text{T},q}^{(n)}}{\chi_{\text{T},q}^{(n)}} }}\right ) \! + \!\sum _{q=1}^{Q}\frac { \frac {\beta_{\text{T},q}^{(n)}}{\chi_{\text{T},q}^{(n)}}}{\ln2(1\!+\! \frac {\beta_{\text{T},q}^{(n)}}{\chi_{\text{T},q}^{(n)}})} \left ({{2\! -\! \frac {\beta_{\text{T},q}^{(n)}}{\beta_{\text{T},q}} \!-\! \frac {\chi_{\text{T},q}}{\chi_{\text{T},q}^{(n)}} }}\right ),\label{21}
\end{aligned}
\end{equation}
where $\beta_{\text{T},q}^{(n)}$ and $\chi_{\text{T},q}^{(n)}$  are the values of $\beta_{\text{T},q}$ and $\chi_{\text{T},q}$ obtained in last iteration, respectively.

Next, we handle the term $\sum\limits_{k = 1}^K {\log_2 (1+\gamma_{\text{R},k})} $ which can be expressed as 
$\sum_{k = 1}^K \log_2 \left(1+\frac{\beta_{\text{R},k}}{\chi _{\text{R},k}}\right)\label{sumk},$
where $\beta_{\text{R},k} = |\mathbf{F}_{\text{R},k} \mathbf{w}_{\text{R},k}|^2$, and a unified notation is adopted for all \( k \). $\chi _{\text{R},k} ={ {\sum\nolimits_{j = k+1}^K |\mathbf{F}_{\text{R},k} \mathbf{w}_{\text{R},j}|^2}+ {\sigma _{\text{R},k}^2}}, \forall k= 1, 2, \dots, K-1 $, and $\chi _{\text{R},K} ={{\sigma _{\text{R},K}^2}}$. Then, we extend $\beta_{\text{R},k}$ as
\vspace{0.01em}
\begin{equation} \small
\begin{aligned}
 \beta&_{\text{R},k}=( {\mathbf{h}}_{\text{R},k}^H {\mathbf{\Phi}}_{\text{b}}^H {\mathbf{G}} + 
\mathbf{h}_{\text{b},k} )\mathbf w_{\text{R},k} \mathbf w_{\text{R},k}^H
(  {\mathbf{G}}^H{\mathbf{\Phi}}_{\text{b}} {\mathbf{h}}_{\text{R},k} + 
\mathbf{h}_{\text{b},k}^H )\\
 =&\mathbf{h}_{\text{b},k}\mathbf w_{\text{R},k} \mathbf w_{\text{R},k}^H
{\mathbf{G}}^H{\mathbf{\Phi}}_{\text{b}} {\mathbf{h}}_{\text{R},k} + 
  {\mathbf{h}}_{\text{R},k}^H {\mathbf{\Phi}}_{\text{b}}^H {\mathbf{G}}\mathbf w_{\text{R},k} \mathbf w_{\text{R},k}^H
{\mathbf{G}}^H
{\mathbf{\Phi}}_{\text{b}} {\mathbf{h}}_{\text{R},k}\\&+
{\mathbf{h}}_{\text{R},k}^H {\mathbf{\Phi}}_{\text{b}}^H {\mathbf{G}}\mathbf w_{\text{R},k} \mathbf w_{\text{R},k}^H\mathbf{h}_{\text{b},k}^H
+\mathbf{h}_{\text{b},k}\mathbf w_{\text{R},k} \mathbf w_{\text{R},k}^H\mathbf{h}_{\text{b},k}^H
\label{23}.
\end{aligned}
\end{equation}

Since all four extended terms in \eqref{23} are scalars and thus equal to their respective traces, the expression can be equivalently rewritten as a summation of trace terms as follows:
\begin{equation} \small
\begin{aligned}
&\beta_{\text{R},k}= \text{Tr}(  {\mathbf{h}}_{\text{R},k}^H {\mathbf{\Phi}}_{\text{b}}^H {\mathbf{G}}\mathbf w_{\text{R},k} \mathbf w_{\text{R},k}^H
{\mathbf{G}}^H
{\mathbf{\Phi}}_{\text{b}} {\mathbf{h}}_{\text{R},k})+
\text{Tr}(\mathbf{h}_{\text{b},k}\mathbf w_{\text{R},k}\\& \mathbf w_{\text{R},k}^H
{\mathbf{G}}^H{\mathbf{\Phi}}_{\text{b}} {\mathbf{h}}_{\text{R},k}) +
\text{Tr}({\mathbf{h}}_{\text{R},k}^H {\mathbf{\Phi}}_{\text{b}}^H {\mathbf{G}}\mathbf w_{\text{R},k} \mathbf w_{\text{R},k}^H\mathbf{h}_{\text{b},k}^H)
+
\text{Tr}(\mathbf{h}_{\text{b},k}\\&\mathbf w_{\text{R},k} \mathbf w_{\text{R},k}^H\mathbf{h}_{\text{b},k}^H)
=\text{Tr}({{\mathbf{\Phi}}_b}\mathbf{X}_{1,k})+\text{Tr}({{\mathbf{\Phi}}_b^H}\mathbf{Y}_{1,k}{\mathbf{\Phi}}_b\mathbf{Z})+
\text{Tr}({{\mathbf{\Phi}}_b}^H\\&\mathbf{X}_{1,k}^H)+\text{Tr}(\mathbf{D}_{1,k}),\label{24}
\end{aligned}
\end{equation}
where $\mathbf{X}_{1,k}={\mathbf{h}}_{\text{R},k}\mathbf{h}_{\text{b},k}\mathbf w_{\text{R},k} \mathbf w_{\text{R},k}^H{\mathbf{G}}^H$, 
$\mathbf{Y}_{1,k}={\mathbf{G}}\mathbf w_{\text{R},k} \mathbf w_{\text{R},k}^H{\mathbf{G}}^H$,
$\mathbf{Z}={\mathbf{h}}_{\text{R},k}{\mathbf{h}}_{\text{R},k}^H$ and
$\mathbf{D}_{1,k}=\mathbf{h}_{\text{b},k}\mathbf w_{\text{R},k} \mathbf w_{\text{R},k}^H\mathbf{h}_{\text{b},k}^H$. As in the previous case, only the beamforming vector \( \mathbf{w}_{\text{R},k} \) varies with the subscript \(k\), while the core structure of each matrix expression remains fixed. 
By using the matrix operation principle in [\cite{zhang2017matrix}, (1.10.6)], the above expression of $\beta_{\text{R},k}$ can be rewritten as
\begin{equation} \small
\begin{aligned}
\beta_{\text{R},k}=&\mathbf{v}_2^T\text{diag}\{\mathbf{X}_{1,k}\} + \mathbf{v}_2^H (\mathbf{Y}_{1,k}\odot\mathbf{Z}^T)\mathbf{v}_2+\text{diag}\{\mathbf{X}_{1,k}^H\}\mathbf{v}_2^{*}+\\\text{Tr}(\mathbf{D}_{1,k})&
=\mathbf{v}_2^H (\mathbf{Y}_{1,k}\odot\mathbf{Z}^T)\mathbf{v}_2 + 2\Re\{\mathbf{v}_2^T\text{diag}\{\mathbf{X}_{1,k}\}\}+\text{Tr}(\mathbf{D}_{1,k}).\label{25}
\end{aligned}
\end{equation}
It can be observed that 
$\beta_{\text{R},k}$ consists of a convex function added to an affine function, and hence remains convex. Similar to the case in \eqref{t1}, we apply a first-order Taylor expansion to approximate the term $ \mathbf{v}_2^H (\mathbf{Y}_{1,k}\odot\mathbf{Z}^T)\mathbf{v}_2$.
\begin{equation} \small\begin{aligned}
 \mathbf{v}_2^H (\mathbf{Y}_{1,k}\odot\mathbf{Z}^T)\mathbf{v}_2\approx& 2\Re{\{{\left({{\mathbf{v}}_\mathrm{{2}}^{(t)}} \right)^{H}}(\mathbf{Y}_{1,k}\odot\mathbf{Z}^T)\}}{{\mathbf{v}}_\mathrm{{2}}}
 \\&+ {\left({{\mathbf{v}}_\mathrm{{2}}^{(t)}} \right)^{H}}(\mathbf{Y}_{1,k}\odot\mathbf{Z}^T){{\mathbf{v}}_\mathrm{{2}}^{(t)}}.
\end{aligned}\end{equation}
where \( \hat{\mathbf{v}}_2^{(t)} \) denotes the feasible point at iteration \( t \), which is updated based on the latest solution. Thus, \( \beta_{\text{R},k} \) can be reformulated as

\begin{equation} \small\begin{aligned}
 \beta_{\text{R},k}&\approx 2\Re{\{{\left({{\mathbf{v}}_\mathrm{{2}}^{(t)}} \right)^{H}}(\mathbf{Y}_{1,k}\odot\mathbf{Z}^T)\}}{{\mathbf{v}}_\mathrm{{2}}}
 + {\left({{\mathbf{v}}_\mathrm{{2}}^{(t)}} \right)^{H}}(\mathbf{Y}_{1,k}\odot\mathbf{Z}^T){{\mathbf{v}}_\mathrm{{2}}^{(t)}}\\&+ 2\Re\{\mathbf{v}_2^T\text{diag}\{\mathbf{X}_{1,k}\}\}+\text{Tr}(\mathbf{D}_{1,k}).
\end{aligned}\end{equation}

We now derive the expression for $\chi_{\text{R},k}$ for $k = 1, 2, \dots, K{-}1$, since $\chi_{\text{R},K}$ is defined as a constant and does not require further processing.
Similarly, for \( \chi_{\text{R},k} \), we define the following intermediate matrices:
\[{\small
\begin{aligned}
\mathbf{X}_{2,j} &= \mathbf{h}_{\text{R},k} \mathbf{h}_{\text{b},k} \mathbf{w}_{\text{R},j} \mathbf{w}_{\text{R},j}^H {\mathbf{G}}^H, \quad
\mathbf{X}_{3,q} = \mathbf{h}_{\text{R},k} \mathbf{h}_{\text{b},k} \mathbf{w}_{\text{T},q} \mathbf{w}_{\text{T},q}^H {\mathbf{G}}^H, \\
\mathbf{Y}_{2,j} &= {\mathbf{G}} \mathbf{w}_{\text{R},j} \mathbf{w}_{\text{R},j}^H {\mathbf{G}}^H, \quad
\mathbf{Y}_{3,q} = {\mathbf{G}} \mathbf{w}_{\text{T},q} \mathbf{w}_{\text{T},q}^H {\mathbf{G}}^H, \\
\mathbf{D}_{2,j} &=  \mathbf{h}_{\text{b},k} \mathbf{w}_{\text{R},j} \mathbf{w}_{\text{R},j}^H \mathbf{h}_{\text{b},k}^H , \quad
\mathbf{D}_{3,q} =   \mathbf{h}_{\text{b},k} \mathbf{w}_{\text{T},q} \mathbf{w}_{\text{T},q}^H \mathbf{h}_{\text{b},k}^H,
\end{aligned}}
\]
thus, the expression for \( \chi_{\text{R},k} \) can be reformulated as
\begin{equation} \small
\begin{aligned}
\chi_{\text{R},k} &= \sum\limits_{j \neq k}^K\mathbf{v}_2^H (\mathbf{Y}_{2,j}\odot\mathbf{Z}^T)\mathbf{v}_2 + \sum\limits_{j \neq k}^K2\Re\{\mathbf{v}_2^T\text{diag}\{\mathbf{X}_{2,j}\}\}\\&+\sum\limits_{j \neq k}^K\text{Tr}(\mathbf{D}_{2,j})+\sum\limits_{q=1}^Q\mathbf{v}_2^H (\mathbf{Y}_{3,q}\odot\mathbf{Z}^T)\mathbf{v}_2 +\sum\limits_{q=1}^Q\text{Tr}(\mathbf{D}_{3,q})\\&+\sum\limits_{q=1}^Q 2\Re\{\mathbf{v}_2^T\text{diag}\{\mathbf{X}_{3,q}\}\} + {\sigma _{\text{R},k}^2}.\label{26}
\end{aligned}
\end{equation}
\vspace{0.01em}
According to \eqref{budengshi}, we can rewrite the term $\sum\nolimits_{k = 1}^K {\log_2 (1+\gamma_{\text{R},k})} $  as
\vspace{-1.0ex}
\begin{equation} \small\begin{aligned}
\sum _{k=1}^{K}\log_2 \left ({{1 + \frac {\beta_{\text{R},k}^{(n)}}{\chi_{\text{R},k}^{(n)}} }}\right )
 + \sum _{k=1}^{K}\frac { \frac {\beta_{\text{R},k}^{(n)}}{\chi_{\text{R},k}^{(n)}}}{\ln2(1+ \frac {\beta_{\text{R},k}^{(n)}}{\chi_{\text{R},k}^{(n)}})} \left ({{2 - \frac {\beta_{\text{R},k}^{(n)}}{\beta_{\text{R},k}} - \frac {\chi_{\text{R},k}}{\chi_{\text{R},k}^{(n)}} }}\right ).\label{27}
\end{aligned}\end{equation}
where $\beta_{\text{R},k}^{(n)}$ and $\chi_{\text{R},k}^{(n)}$ are the values of $\beta_{\text{R},k}$ and $\chi_{\text{R},k}$ obtained in last iteration, respectively. Therefore, by combining \eqref{21} and \eqref{27}, the objective function can be reformulated as
\begin{equation} \small\begin{aligned}
&\sum _{q=1}^{Q}\log_2 \left ({{1 \! +\! \frac {\beta_{\text{T},q}^{(n)}}{\chi_{\text{T},q}^{(n)}} }}\right ) \! + \!\sum _{q=1}^{Q}\frac { \frac {\beta_{\text{T},q}^{(n)}}{\chi_{\text{T},q}^{(n)}}}{\ln2(1\!+\! \frac {\beta_{\text{T},q}^{(n)}}{\chi_{\text{T},q}^{(n)}})} \left ({{2\! -\! \frac {\beta_{\text{T},q}^{(n)}}{\beta_{\text{T},q}} \!-\! \frac {\chi_{\text{T},q}}{\chi_{\text{T},q}^{(n)}} }}\right )+\\&\sum _{k=1}^{K}\log_2 \left ({{1 + \frac {\beta_{\text{R},k}^{(n)}}{\chi_{\text{R},k}^{(n)}} }}\right )
 + \sum _{k=1}^{K}\frac { \frac {\beta_{\text{R},k}^{(n)}}{\chi_{\text{R},k}^{(n)}}}{\ln2(1+ \frac {\beta_{\text{R},k}^{(n)}}{\chi_{\text{R},k}^{(n)}})} \left ({{2 - \frac {\beta_{\text{R},k}^{(n)}}{\beta_{\text{R},k}} - \frac {\chi_{\text{R},k}}{\chi_{\text{R},k}^{(n)}} }}\right ).\label{28}
\end{aligned}\end{equation}

Then, constraints \eqref{P1d}, \eqref{P1e}, and \eqref{P1f}, which exhibit the same structural pattern as the expressions derived above, can be reformulated accordingly. Specifically, they can be rewritten as{\small
\begin{align}\small
 &\frac {\beta_{\text{T},q}}{\chi_{\text{T},q}} \geq \gamma_{\text{min}}, \quad \forall q = 1, 2, \dots, Q,\label{37}\\
& \frac {\beta_{\text{R},k}}{\chi_{\text{R},k}} \geq \gamma_{\text{min}},\quad \forall k = 1, 2, \dots, K, \label{40}\end{align}}
\vspace{0.02em}
 \begin{equation} \small\begin{aligned}
& 2\Re{\{{\left({{\mathbf{v}}_\mathrm{{1}}^{(t)}} \right)^{H}}{\mathbf{U}_{3,K}}\}}{{\mathbf{v}}_\mathrm{{1}}}+ {\left({{\mathbf{v}}_\mathrm{{1}}^{(t)}} \right)^{H}}{\mathbf{U}_{3,K}}{{\mathbf{v}}_\mathrm{{1}}^{(t)}}\leq \cdots \leq\\& 2\Re{\{{\left({{\mathbf{v}}_\mathrm{{1}}^{(t)}} \right)^{H}}{\mathbf{U}_{1,1}}\}}{{\mathbf{v}}_\mathrm{{1}}}+ {\left({{\mathbf{v}}_\mathrm{{1}}^{(t)}} \right)^{H}}{\mathbf{U}_{1,1}}{{\mathbf{v}}_\mathrm{{1}}^{(t)}}^H,\label{41}
\end{aligned}\end{equation}
\begin{equation} \small\begin{aligned}
&2\Re{\{{\left({{\mathbf{v}}_\mathrm{{2}}^{(t)}} \right)^{H}}(\mathbf{Y}_{1,K}\odot\mathbf{Z}^T)\}}{{\mathbf{v}}_\mathrm{{2}}}
 + {\left({{\mathbf{v}}_\mathrm{{2}}^{(t)}} \right)^{H}}(\mathbf{Y}_{1,K}\odot\mathbf{Z}^T){{\mathbf{v}}_\mathrm{{2}}^{(t)}}\\&+ 2\Re\{\mathbf{v}_2^T\text{diag}\{\mathbf{X}_{1,K}\}\}+\text{Tr}(\mathbf{D}_{1,K})\leq \cdots \leq \\&2\Re{\{{\left({{\mathbf{v}}_\mathrm{{2}}^{(t)}} \right)^{H}}(\mathbf{Y}_{3,1}\odot\mathbf{Z}^T)\}}{{\mathbf{v}}_\mathrm{{2}}}
 + {\left({{\mathbf{v}}_\mathrm{{2}}^{(t)}} \right)^{H}}(\mathbf{Y}_{3,1}\odot\mathbf{Z}^T){{\mathbf{v}}_\mathrm{{2}}^{(t)}}+ \\&2\Re\{\mathbf{v}_2^T\text{diag}\{\mathbf{X}_{3,1}\}\}+\text{Tr}(\mathbf{D}_{3,1}).\label{P42}
 \end{aligned}\end{equation}
 
Therefore, we rewrite the sub-problem as
\vspace*{-0.2em}{\small
\begin{align}\small
\mathop {\max }\limits_{\mathbf{v}_1,\mathbf{v}_2}&\ \ \eqref{28}\label{P2_OF2}\\
{\text{s.t.}}\ \ & \ \ \mathbf{v}_1(m)\mathbf{v}_1^H(m)+ \mathbf{v}_2(m)\mathbf{v}_2^H(m)\leq 1, \forall m \in M, \tag{\ref{P2_OF2}{a}} \\
\ \ & \ \ \eqref{37}-\eqref{P42} \tag{\ref{P2_OF2}{b}}\label{P22a} ,
\end{align}}
which is convex, and the corresponding optimal solutions can be obtained by CVX.
\vspace*{-0.5em}
\subsection{Optimization of the FIRES Element Positions}
In this subsection, we optimize the FIRES element positions \( \mathbf{p} \), assuming fixed beamforming vectors \( \{ \mathbf{w}_{\text{R},k} \}_{k=1}^{K} \), \( \{ \mathbf{w}_{\text{T},q} \}_{q=1}^{Q} \), and FIRES coefficients \( \boldsymbol{\Phi}_{\text{b}} \) and \( \boldsymbol{\Theta}_{\text{b}} \). Introducing auxiliary variables \( \Xi_k \) and \( \Upsilon_q \), and omitting terms independent of \( \mathbf{p} \), the sub-problem is expressed as{\small
\begin{align}\small
&\mathop {\max }\limits_{\mathbf{p},\{\Xi_k\}_{k=1}^{K},\{\Upsilon_q\}_{q=1}^{Q}} \quad  \sum\limits_{k = 1}^K \log_2 (1+\Xi_k) + \sum\limits_{q = 1}^Q \log_2 (1+\Upsilon_q)  \label{P3_OF}\\
\text{s.t.} \quad & \gamma_{\text{R},k} \geq \Xi_k , \quad \forall k=1, 2, \dots,K, \tag{\ref{P3_OF}{a}}\label{P3a} \\
& \gamma_{\text{T},q} \geq \Upsilon_q , \quad \forall q=1, 2, \dots,Q, \tag{\ref{P3_OF}{b}}\label{P3b} \\
& \eqref{P1b}-\eqref{P1f}, \tag{\ref{P3_OF}{c}} \label{P3c} 
\end{align}}

As outlined in Section II-A, the channels \( \mathbf{h}_{\text{T},q} \), \( \mathbf{h}_{\text{R},k} \), and \( \mathbf{h}_{\text{b},k} \) are functions of the FIRES element positions. The optimization problem in \eqref{P3_OF} presents significant challenges, primarily due to the non-convex constraints. These constraints exhibit a highly intricate structure with respect to the position variable \( \mathbf{p} \). Specifically, the dependency of the channels on \( \mathbf{p} \) introduces mixed quadratic and quartic terms in the outer function. Moreover, the inner structure of the constraints involves complex exponential terms arising from the array response, as defined in \eqref{chuchuwen}, which further increases the mathematical complexity of the problem.

 We begin by focusing on the constraint~\eqref{P3a}. We define the compact notations 
$\mathcal{C}_{\text{R},k} = |\mathbf{F}_{\text{R},k} \mathbf{w}_{\text{R},k}|^2$, which holds uniformly for all $k$. 
For the denominator, we define $\mathcal{D}_{\text{R},k} = \sum_{j =k+1}^{K} |\mathbf{F}_{\text{R},k} \mathbf{w}_{\text{R},j}|^2+ {\sigma _{\text{R},k}^2}$ for $k = 1, 2, \dots, K{-}1$, while for $k = K$, $\mathcal{D}_{\text{R},K} = {\sigma _{\text{R},K}^2}$ becomes a constant. 
The non-convexity of the constraint arises from the fact that both the numerator and denominator generally depend on the optimization variables, and the presence of the variable $\Xi_k$ on the right-hand side introduces additional bilinearity.
 Thus, we reformulate the constraint using the Dinkelbach method \cite{8314727} as
 \vspace{0.02em}
\begin{equation} \small
\begin{aligned}
\mathcal{C}_{\text{R},k} - y_{\text{R},k}^{(n)}
\mathcal{D}_{\text{R},k}\geq \Xi_k,\forall k = 1, 2, \cdots,K,\label{34}
\end{aligned}
\end{equation}

\noindent where  $y_{\text{R},k}^{(n)}$ denotes the value of $y_{\text{R},k}$ at the $n$th iteration, which is defined as $y_{\text{R},k}^{(n)} =\mathcal{C}_{\text{R},k} /\left(\mathcal{D}_{\text{R},k}+ {\sigma _{\text{R},k}^2}\right)^{(n-1)} $. \eqref{34} is still non-convex due to the  highly intricate structures of $\mathcal{C}_{\text{R},k}$ and $\mathcal{D}_{\text{R},k}$. To emphasize the dependency on the position variable \( \mathbf{p} \), we omit terms that are independent of \( \mathbf{p} \), and rewrite the expressions of \( \mathcal{C}_{\text{R},k} \) and \( \mathcal{D}_{\text{R},k} \) accordingly. It can be observed that \( \mathcal{C}_{\text{R},k} \) and \( \mathcal{D}_{\text{R},k} \) share the same structural form, differing only in the beamforming vectors \( \mathbf{w}_{\text{R},k} \), \( \mathbf{w}_{\text{R},j} \), and \( \mathbf{w}_{\text{T},q} \), while all other components remain unchanged. Therefore, given the effective channel
$\mathbf{F}_{\text{R},k} = {\mathbf{h}}_{\text{R},k}^H {\mathbf{\Phi}}_{\text{b}}^H {\mathbf{G}} + \mathbf{h}_{\text{b},k},$ all relevant terms can be uniformly expressed as
 \begin{equation} \small
\begin{aligned}
&|\mathbf{F}_{\text{R},k} \mathbf{w}_\mu|^2= \mathbf{A}_{\mu,\text{1}}\!+\!\mathbf{A}_{\mu,\text{2}}\!+\!\mathbf{A}_{\mu,\text{3}}\!+\!\mathbf{A}_{\mu,\text{4}}
=Z_{\mu,\text{1}}\mathbf{a}_{\text{r}}(\mathbf{p})^H
\overline{\mathbf{a}}_{\text{R},k}(\mathbf{p})\mathbf{v}_2\\&\mathbf{v}_2^H\overline{\mathbf{a}}_{\text{R},k}(\mathbf{p})^H\mathbf{a}_{\text{r}}(\mathbf{p})\!+\!Z_{\mu,\text{2}}\overline{\mathbf{a}}_{r}(\mathbf{p})\mathbf{a}_{\text{R},k}( \mathbf{p})\!\!+\!\!Z_{\mu,\text{3}}\mathbf{a}_{\text{R},k}^H( \mathbf{p})\overline{\mathbf{a}}_{r}^H(\mathbf{p})\!+\!\mathbf{A}_{\mu,\text{4}},\label{A-a}
\end{aligned}
\end{equation}
where \( \mu \in \{ (\text{R},k), (\text{R},j), (\text{T},q) \} \) is a unified beamforming index, and \( \mathbf{A}_{\mu,4} = \mathbf{w}_\mu^H \mathbf{h}_{\text{b},k}^H \mathbf{h}_{\text{b},k} \mathbf{w}_\mu \), which is independent of \( \mathbf{p} \). Detailed derivations and notations are provided in Appendix~A.

Accordingly, we obtain
 \begin{equation} \small
\begin{aligned}
\mathcal{C}_{\text{R},k}= \mathbf{A}_{\text{R},k,\text{1}}+\mathbf{A}_{\text{R},k,\text{2}}+\mathbf{A}_{\text{R},k,\text{3}}+\mathbf{A}_{\text{R},k,\text{4}}
,\label{ck1}
\end{aligned}
\end{equation}
 \begin{equation} \small
\begin{aligned}
\mathcal{D}_{\text{R},k}= &\sum_{j =k+1}^{K}(\mathbf{A}_{\text{R},j,\text{1}}+\mathbf{A}_{\text{R},j,\text{2}}+\mathbf{A}_{\text{R},j,\text{3}}+\mathbf{A}_{\text{R},j,\text{4}})+ {\sigma _{\text{R},k}^2}
.\label{dk1}
\end{aligned}
\end{equation}
 However, \eqref{ck1} and \eqref{dk1} still depend on the array manifold, which introduces significant challenges for optimization. Therefore, we further simplify them by optimizing the position of one FIRES element at a time, i.e., by focusing on the position of the \( l \)th element \( \mathbf{p}_l \), while keeping all other elements fixed. The corresponding expression with respect to the position of the \( l \)th element \( \mathbf{p}_l \) is given by
 \begin{equation} \small
\begin{aligned}
\mathcal{C}_{\text{R},k}(\mathbf{p}_l)= \mathbf{A}_{\text{R},k,\text{1}}(\mathbf{p}_l)+\mathbf{T}_{\text{R},k}(\mathbf{p}_l)+\mathbf{A}_{\text{R},k,\text{4}}
,\label{ck2}
\end{aligned}
\end{equation}
 \begin{equation} \small
\begin{aligned}
\mathcal{D}_{\text{R},k}(\mathbf{p}_l)= &\sum_{j = k+1}^{K}(\mathbf{A}_{\text{R},j,\text{1}}(\mathbf{p}_l)+\mathbf{T}_{\text{R},j}(\mathbf{p}_l)+\mathbf{A}_{\text{R},j,\text{4}})+ {\sigma _{\text{R},k}^2}
.\label{dk2}
\end{aligned}
\end{equation}
 Detailed derivations and symbol explanations are provided in Appendix~B.
 
However, the expressions in \eqref{ck2} and \eqref{dk2} involve both cosine and sine terms, which are non-convex with respect to the position variable \( \mathbf{p}_l \), making the optimization problem intractable in its original form. 

To approximate the non-convex terms in the constraint functions, we adopt the MM framework. Specifically, we construct either concave minorizers or convex majorizers using the second-order Taylor expansion, depending on whether the corresponding term is being maximized or appears in a constraint.

For a twice-differentiable function \( f : \mathbb{R}^n \to \mathbb{R} \), its second-order Taylor expansion around a given point \( \mathbf{x}^{(n)} \in \mathbb{R}^n \)is given by
\begin{equation} \small
\begin{aligned}
f(\mathbf{x}) \approx f(\mathbf{x}^{(n)})& + \nabla f(\mathbf{x}^{(n)})^T (\mathbf{x} - \mathbf{x}^{(n)})\\& + \frac{1}{2} (\mathbf{x} - \mathbf{x}^{(n)})^T \nabla^2 f(\mathbf{x}^{(n)}) (\mathbf{x} - \mathbf{x}^{(n)}),
\label{eq:taylor2}
\end{aligned}
\end{equation}
where \( \nabla f(\mathbf{x}^{(n)}) \in \mathbb{R}^n \) denotes the gradient vector and \( \nabla^2 f(\mathbf{x}^{(n)}) \in \mathbb{R}^{n \times n} \) denotes the Hessian matrix of \( f \) evaluated at \( \mathbf{x}^{(n)}\).
Then, by upper-bounding the Hessian via \( \nabla^2 f(\mathbf{x}^{(n)}) \preceq \delta \mathbf{I} \), we obtain the following concave lower bound and convex upper bound, which can be used in optimization:

\begin{lemma}\small \small
\label{lemma:taylor}
Let \( f(\mathbf{x}) \) be a twice-differentiable function. Then, around \( \mathbf{x}^{(n)} \), the following inequality holds:
\[
f(\mathbf{x}) \geq f(\mathbf{x}^{(n)}) + \nabla f(\mathbf{x}^{(n)})^T(\mathbf{x} - \mathbf{x}^{(n)}) - \frac{\delta}{2} \| \mathbf{x} - \mathbf{x}^{(n)} \|^2,
\]
\[
f(\mathbf{x}) \leq f(\mathbf{x}^{(n)}) + \nabla f(\mathbf{x}^{(n)})^T(\mathbf{x} - \mathbf{x}^{(n)}) + \frac{\delta}{2} \| \mathbf{x} - \mathbf{x}^{(n)} \|^2,
\]
where \( \delta \in \mathbb{R}_{+} \) satisfies \( \nabla^2 f(\mathbf{x}^{(n)}) \preceq \delta \mathbf{I} \). A conservative yet valid choice for \( \delta \) is to set
\[
\delta = \left\| \nabla^2 f(\mathbf{x}^{(n)}) \right\|_F,
\]
where \( \| \cdot \|_F \) denotes the Frobenius norm. This choice guarantees the majorization condition, although it may not yield the tightest bound.

\end{lemma}

Therefore, by applying Lemma 1, we obtain a lower-bound approximation for \eqref{ck2} and an upper-bound approximation for \eqref{dk2}, respectively. Following the derivation of gradient and Hessian matrix in Appendix C, we define $\nabla \tilde{f}_{\mu}({\mathbf{p}_l}) = \nabla \mathbf{A}_{\mu}({\mathbf{p}_l})+\nabla \mathbf{T}_{\mu}({\mathbf{p}_l})$ and $\nabla^2 \tilde{f}_{\mu}({\mathbf{p}_l})=\nabla^2 \mathbf{A}_{\mu}({\mathbf{p}_l})+\nabla^2 \mathbf{T}_{\mu}({\mathbf{p}_l})$, where \( \mu \in \{ (\text{R},k), (\text{R},j), (\text{T},q) \} \). Hence, it follows that $\delta_{\mu}^{\text{R}} = \left\| \nabla^2 \tilde{f}_{\mu}({\mathbf{p}_l}) \right\|_F$. The resulting lower-bound expression of \eqref{ck2} is given by:
\begin{equation} \small
\begin{aligned}
\tilde {\mathcal{C}}_{\text{R},k}(\mathbf{p}_l) \!= {\mathcal{C}_{\text{R},k}}(\mathbf{p}_l^{(t)})+ \nabla \tilde{f}_{\text{R},k}(\mathbf{p}_l^{(t)})^T
\!(\mathbf{p}_l \!-\! \mathbf{p}_l^{(t)}) \!-\!\frac{\delta^{\text{R}}_{\text{R},k}}{2}\|\mathbf{p}_l - \mathbf{p}_l^{(t)}\|^2 ,\label{41}
\end{aligned}
\end{equation}
and the resulting upper-bound expression of \eqref{dk2} is given by:
\begin{equation} \small
\begin{aligned}
\tilde {\mathcal{D}}_k&(\mathbf{p}_l) = {\mathcal{D}_{\text{R},k}}(\mathbf{p}_l^{(t)})\!\!\!+ \!\!\!\sum_{j =k+1}^{K}\nabla \tilde{f}_{\text{R},j}(\mathbf{p}_l^{(t)})^T
(\mathbf{p}_l - \mathbf{p}_l^{(t)})\\&+\frac{\delta^{\text{R}}_{\text{R},j}}{2}\|\mathbf{p}_l \!\!-\! \mathbf{p}_l^{(t)}\|^2+ {\sigma _{\text{R},k}^2}, \label{42}
\end{aligned}
\end{equation}
where $\mathbf{p}_l^{(t)}$ represents a feasible point at the $t$th iteration. Note that after each iteration, $\mathbf{p}_l^{(t)}$ will be updated according to the optimal solutions obtained. 

Therefore, we can rewrite the constraint \eqref{34} as
\begin{equation} \small
\begin{aligned}
\tilde {\mathcal{C}}_{\text{R},k}(\mathbf{p}_l) - y_{\text{R},k}^{(n)}
\left(\tilde {\mathcal{D}}_k(\mathbf{p}_l)+ {\sigma _{\text{R},k}^2}\right)\geq \Xi_k,\forall k=1, 2, \cdots,K,\label{43}
\end{aligned}
\end{equation}
where \( \tilde{\mathcal{C}}_k(\mathbf{p}_l) \) is concave and \( \tilde{\mathcal{D}}_k(\mathbf{p}_l) \) is convex, their difference in \eqref{43} remains concave (or constant when \( k=K \)). Hence, \eqref{43} defines a convex constraint for all \( k \).

Then, we address the non-convexity of constraint \eqref{P3b}. Similarly, we define the notations $\mathcal{C}_{\text{T},q}  = |\mathbf{F}_{\text{T},q}\mathbf{w}_{\text{T},q}|^2$, which holds uniformly for all $q$. $\mathcal{D}_{\text{T},q}=\sum\nolimits_{j= q+1}^{Q} |\mathbf{F}_{\text{T},q} \mathbf{w}_{\text{T},j}|^2 + \sum\nolimits_{k=1}^{K} |\mathbf{F}_{\text{T},q} \mathbf{w}_{\text{R},k}|^2+ {\sigma _{\text{T},q}^2}, \forall q= 1, \cdots, Q-1 $, and    $\mathcal{D}_{\text{T},Q}=\sum\nolimits_{k=1}^{K} |\mathbf{F}_{\text{T},Q} \mathbf{w}_{\text{R},Q}|^2+ {\sigma _{\text{T},Q}^2}$. By employing the Dinkelbach method, the constraint can be equivalently reformulated as
\begin{equation} \small
\begin{aligned}
\mathcal{C}_{\text{T},q} - y_{\text{T},q}^{(n)}
\left(\mathcal{D}_{\text{T},q}+ {\sigma _{\text{T},q}^2}\right)\geq \Upsilon_q,\forall q=1, 2, \cdots,Q,\label{56}
\end{aligned}
\end{equation}
where  $y_{\text{T},q}^{(n)}$ denotes the value of $y_{\text{T},q}$ at the $n$th iteration, which is defined as $y_{\text{T},q}^{(n)} =\mathcal{C}_{\text{T},q}/\left(\mathcal{D}_{\text{T},q}+ {\sigma _{\text{T},q}^2}\right)^{(n-1)} $. 

The position of each FIRES element is optimized in a sequential manner. To facilitate this process, the MM framework is employed. Detailed derivations and notations are provided in Appendix~D. Accordingly, we define \( \delta_{\mu}^{\text{T}} = \left\| \nabla^2 \mathbf{B}_{\mu}(\mathbf{p}_l) \right\|_F \), where \( \mu \in \{ (\text{R},k), (\text{R},j), (\text{T},q) \} \). Based on the constructed surrogate functions under the MM framework, the lower bound of \( \mathcal{C}_{\text{T},q}(\mathbf{p}_l) \) and the upper bound of \( \mathcal{D}_{\text{T},q}(\mathbf{p}_l) \) can be reformulated as
\vspace{-1.0ex}
\begin{equation} \small 
\begin{aligned}
&\tilde {\mathcal{C}}_{\text{T},q}(\mathbf{p}_l) = {\mathcal{C}_{\text{T},q}}(\mathbf{p}_l^{(t)})+ \nabla \mathbf{B}_{\text{T},q}(\mathbf{p}_l^{(t)})^T
(\mathbf{p}_l - \mathbf{p}_l^{(t)}) \\&-\frac{\delta_{\text{T},q}^{\text{T}}}{2}\|\mathbf{p}_l - \mathbf{p}_l^{(t)}\|^2 , \,\,\,\,\,\,\,\,\,\,\,\,\,\,\,\,\,\,\,\,\,\,\,\,\,\,\,\,\,\,\,\,\,\,\,\,\,\,\,\,\forall q=1, 2, \cdots,Q,\label{Q_upp}
\end{aligned}
\end{equation}
\begin{equation} \small
\begin{aligned}
\tilde {\mathcal{D}}_{\text{T},q}&(\mathbf{p}_l) = {\mathcal{D}_{\text{T},q}}(\mathbf{p}_l^{(t)})\!+\! \sum_{j=q\!+\!1}^{Q}\nabla  \mathbf{B}_{\text{T},j}(\mathbf{p}_l^{(t)})^T
(\mathbf{p}_l - \mathbf{p}_l^{(t)}) \\&\!+\!\sum_{j =q\!+\!1}^{Q}\frac{\delta^{\text{T}}_{\text{T},j}}{2}\|\mathbf{p}_l - \mathbf{p}_l^{(t)}\|^2\!+\!\sum_{k=1}^{K}\nabla \mathbf{B}_{\text{R},k}(\mathbf{p}_l^{(t)})^T
(\mathbf{p}_l - \mathbf{p}_l^{(t)})\\&\!+\!\sum_{k=1}^{K}\frac{\delta^{\text{T}}_{\text{R},k}}{2}\|\mathbf{p}_l - \mathbf{p}_l^{(t)}\|^2 , \,\,\,\,\,\,\,\,\forall q=1, 2, \cdots,Q-1,
\end{aligned}
\end{equation}
\begin{equation} \small
\begin{aligned}
&\tilde {\mathcal{D}}_{\text{T},Q}(\mathbf{p}_l) = {\mathcal{D}_{\text{T},Q}}(\mathbf{p}_l^{(t)})+\sum_{k=1}^{K}\nabla \mathbf{B}_{\text{R},k}(\mathbf{p}_l^{(t)})^T
(\mathbf{p}_l - \mathbf{p}_l^{(t)})\\&+\sum_{k=1}^{K}\frac{\delta^{\text{T}}_{\text{R},k}}{2}\|\mathbf{p}_l - \mathbf{p}_l^{(t)}\|^2,
\end{aligned}
\end{equation}
where $\mathbf{p}_l^{(t)}$ represents a feasible point at the $t$th iteration. Note that after each iteration, $\mathbf{p}_l^{(t)}$ will be updated according to the optimal solutions obtained. Therefore, \eqref{56} can be rewritten as 
\begin{equation} \small
\begin{aligned}
\tilde {\mathcal{C}}_{\text{T},q}(\mathbf{p}_l) - y_{\text{T},q}^{(n)}
\left(\tilde {\mathcal{D}}_{\text{T},q}(\mathbf{p}_l)+ {\sigma _{\text{T},q}^2}\right)\geq \Upsilon_q,\forall q=1, 2, \cdots,Q,\label{P3C}
\end{aligned}
\end{equation}

Next, we handle the constraints \eqref{P1d},\eqref{P1e} and \eqref{P1f}. Note that the above constraints share the same mathematical structure as those reformulated in the previous subsection. Therefore, they can be  reformulated as{\small
\begin{align}\small
 &\frac {\tilde {\mathcal{C}}_{\text{T},q}(\mathbf{p}_l)}{\tilde {\mathcal{D}}_{\text{T},q}(\mathbf{p}_l)} \geq \gamma_{\text{min}}, \quad \forall q = 1, 2, \dots, Q,\\
& \frac {\tilde {\mathcal{C}}_{\text{R},k}(\mathbf{p}_l)}{\tilde {\mathcal{D}}_{\text{R},k}(\mathbf{p}_l)} \geq \gamma_{\text{min}},\quad \forall k = 1, 2, \dots, K, \end{align}}

\begin{equation} \small 
\begin{aligned}
 & \mathbf{E}_{\text{R},K} \leq \cdots \leq \mathbf{E}_{\text{T},1},\\ 
 & \mathbf{S}_{\text{R},K} \leq \cdots \leq \mathbf{S}_{\text{T},1},\label{last}
\end{aligned}
\end{equation}
where \( \mathbf{E}_{\mu} \) and \( \mathbf{S}_{\mu} \) represent the lower-bound approximations of \( \mathcal{C}_{\text{R},k}(\mathbf{p}_l) \) and \( \mathcal{C}_{\text{T},q}(\mathbf{p}_l) \), respectively, as derived in \eqref{41} and \eqref{Q_upp}. The variation in the subscript \( \mu \) corresponds to the change in the associated beamforming vector \( \mathbf{w}_{\mu} \), while the core channel-related terms remain fixed. Since the derivation process is entirely analogous to that presented earlier, it is omitted here for brevity. Notably, all the resulting constraints are convex with respect to \( \mathbf{p}_l \).

In addition, constraint~\eqref{P1b} is still non-convex. Therefore, we approximate it using the first-order Taylor expansion within the SCA method. The resulting approximation is given as follows:
\begin{equation} \small 
\begin{aligned}
\frac{
\left( \mathbf{p}_l^{(t-1)} - \mathbf{p}_{l'} \right)^T \left( \mathbf{p}_l - \mathbf{p}_{l'} \right)
}{
\left\| \mathbf{p}_l^{(t-1)} - \mathbf{p}_{l'} \right\|
} \geq \Delta D, \forall l \ne l' \in \{1, 2, \dots,L\}, \label{last1}
\end{aligned}
\end{equation}
which is convex with respect to \( \mathbf{p}_l \).
Therefore, we can rewrite the sub-problem \eqref{P3_OF} as{\small
\begin{align}\small
&\mathop {\max }\limits_{\mathbf{p,\Xi_k,\Upsilon_q}}\ \ \sum\limits_{k = 1}^K {\log_2 (1+\Xi_k)} +\sum\limits_{q = 1}^Q {\log_2 (1+\Upsilon_q)}  \label{P3_OF1}\\
{\text{s.t.}}\ \ & \ \ \eqref{43}, \eqref{P3C}-\eqref{last1},\tag{\ref{P3_OF1}{a}} 
\end{align}}

which is convex, and the corresponding optimal solutions can be obtained by CVX. The detailed steps of the proposed algorithm are summarized in Algorithm~1. 

\begin{algorithm}[t]\small
\caption{{Successive Optimization of FIRES Element Positions}}
\begin{algorithmic}[1]
    \STATE Initialize $l = 1$ and $\mathbf{p}^{(0)}$.
   \REPEAT 
        \STATE Initialize $i = 0$, feasible $y_{\text{R},k}^{(0)}$, $y_{\text{T},q}^{(0)}$, and tolerance $\epsilon = 10^{-1}$.
         \REPEAT 
        \STATE Solve problem \eqref{P3_OF1} with fixed $y_{\text{R},k}^{(i)}$, $y_{\text{T},q}^{(i)}$ to obtain the position $\mathbf{p}_l^{(i)}$ of the $l$th RIS element.
        \STATE Update $y_{\text{R},k}^{(i+1)}$ and $y_{\text{T},q}^{(i+1)}$ using \eqref{34} and \eqref{56}.
        \STATE Set $i = i + 1$.
    \UNTIL $\frac{y_{\text{R},k}^{(i+1)}-y_{\text{R},k}^{(i)}}{y_{\text{R},k}^{(i)}} < \epsilon$ and $\frac{y_{\text{T},q}^{(i+1)}-y_{\text{T},q}^{(i)}}{y_{\text{T},q}^{(i)}} < \epsilon$.
    \STATE Set $l = l + 1$. 
\UNTIL $l > L$. 
\end{algorithmic}
\end{algorithm}
\begingroup

\subsection{Convergence Analysis}

In this subsection, we analyze the convergence of the proposed Algorithm~2,
which is developed under an AO framework.

Define the collection of transmit beamforming vectors at iteration $t$ as
$\mathcal W^{(t)} \triangleq 
\big\{\{\mathbf w_{\mathrm R,k}^{(t)}\}_{k=1}^{K},
\{\mathbf w_{\mathrm T,q}^{(t)}\}_{q=1}^{Q}\big\}$.
The objective value at the $t$th outer iteration is given by
\begin{equation} \small
\label{eq:R_t}
R^{(t)} \triangleq 
\sum_{k=1}^{K}\log_2\!\left(1+\gamma_{\mathrm R,k}^{(t)}\right)
+\sum_{q=1}^{Q}\log_2\!\left(1+\gamma_{\mathrm T,q}^{(t)}\right).
\end{equation}

\subsubsection*{1) Update of Transmit Beamforming Vectors}

For fixed $\boldsymbol{\Phi}_{\mathrm b}^{(t)}$, $\boldsymbol{\Theta}_{\mathrm b}^{(t)}$,
and $\mathbf p^{(t)}$, the transmit beamforming vectors
$\{\mathbf w_{\mathrm R,k}\}_{k=1}^{K}$ and $\{\mathbf w_{\mathrm T,q}\}_{q=1}^{Q}$
are optimized via SDR and dropping the rank-one
constraints.
Since SDR enlarges the feasible set, the relaxed problem provides an upper bound
on the original objective value \cite{8811733}.

To handle the remaining non-convex fractional logarithmic terms,
inequality~\eqref{budengshi} is applied to construct a tight concave lower bound
around the previous iterate, leading to the convex SDP in~\eqref{P1_OF3}.
Denote the original SDR objective by $F(\boldsymbol{\Omega}_1)$ and the surrogate
objective by $\tilde F(\boldsymbol{\Omega}_1\mid\boldsymbol{\Omega}_1^{(t)})$.
The surrogate satisfies
\begin{equation} \small
\begin{aligned}
\tilde F(\boldsymbol{\Omega}_1\mid\boldsymbol{\Omega}_1^{(t)})
&\le F(\boldsymbol{\Omega}_1),\\
\tilde F(\boldsymbol{\Omega}_1^{(t)}\mid\boldsymbol{\Omega}_1^{(t)})
&= F(\boldsymbol{\Omega}_1^{(t)}).
\end{aligned}
\end{equation}
Since \eqref{P1_OF3} is solved, we have
\begin{equation} \small
F(\boldsymbol{\Omega}_1^{(t+1)})
\ge F(\boldsymbol{\Omega}_1^{(t)}),
\end{equation}
which indicates that the beamforming update yields a non-decreasing objective
sequence. Moreover, by the rank reduction result in~\cite{rank1}, a rank-one
optimal solution exists and the beamforming vectors can be recovered.

\subsubsection*{2) Update of FIRES Coefficients}

With the transmit beamforming vectors and FIRES element positions fixed,
the coefficients $\boldsymbol{\Phi}_{\mathrm b}$ and $\boldsymbol{\Theta}_{\mathrm b}$
are optimized via the SCA framework.

At iteration $t$, all non-convex terms in the objective function are replaced
by tight concave surrogate functions constructed based on
inequality~\eqref{budengshi} and the first-order Taylor expansions derived above.
Denote by $R(\boldsymbol{\Phi}_{\mathrm b},\boldsymbol{\Theta}_{\mathrm b})$
the original objective of problem~\eqref{P2_OF}, and by
$\tilde R(\boldsymbol{\Phi}_{\mathrm b},\boldsymbol{\Theta}_{\mathrm b}
\mid \boldsymbol{\Phi}_{\mathrm b}^{(t)},\boldsymbol{\Theta}_{\mathrm b}^{(t)})$
the corresponding surrogate objective adopted in~\eqref{P2_OF2}.
By construction, the surrogate satisfies
\begin{equation} \small
\begin{aligned}
\tilde R(\boldsymbol{\Phi}_{\mathrm b},\boldsymbol{\Theta}_{\mathrm b}
\mid \boldsymbol{\Phi}_{\mathrm b}^{(t)},\boldsymbol{\Theta}_{\mathrm b}^{(t)})
&\le
R(\boldsymbol{\Phi}_{\mathrm b},\boldsymbol{\Theta}_{\mathrm b}),\\
\tilde R(\boldsymbol{\Phi}_{\mathrm b}^{(t)},\boldsymbol{\Theta}_{\mathrm b}^{(t)}
\mid \boldsymbol{\Phi}_{\mathrm b}^{(t)},\boldsymbol{\Theta}_{\mathrm b}^{(t)})
&=
R(\boldsymbol{\Phi}_{\mathrm b}^{(t)},\boldsymbol{\Theta}_{\mathrm b}^{(t)}).
\end{aligned}
\end{equation}

At each iteration, the convex subproblem~\eqref{P2_OF2} is solved,
which yields
\begin{equation} \small
\tilde R(\boldsymbol{\Phi}_{\mathrm b}^{(t+1)},\boldsymbol{\Theta}_{\mathrm b}^{(t+1)}
\mid \boldsymbol{\Phi}_{\mathrm b}^{(t)},\boldsymbol{\Theta}_{\mathrm b}^{(t)})
\ge
\tilde R(\boldsymbol{\Phi}_{\mathrm b}^{(t)},\boldsymbol{\Theta}_{\mathrm b}^{(t)}
\mid \boldsymbol{\Phi}_{\mathrm b}^{(t)},\boldsymbol{\Theta}_{\mathrm b}^{(t)}).
\end{equation}
Combining the above inequalities leads to
\begin{equation} \small
R(\boldsymbol{\Phi}_{\mathrm b}^{(t+1)},\boldsymbol{\Theta}_{\mathrm b}^{(t+1)})
\ge
R(\boldsymbol{\Phi}_{\mathrm b}^{(t)},\boldsymbol{\Theta}_{\mathrm b}^{(t)}),
\end{equation}
which shows that the objective value is non-decreasing over the SCA iterations.

Moreover, the feasible set of $\boldsymbol{\Phi}_{\mathrm b}$ and
$\boldsymbol{\Theta}_{\mathrm b}$ is compact due to the element-wise amplitude
constraints, and the objective function is upper bounded.
Therefore, the proposed SCA-based FIRES coefficient optimization
is guaranteed to converge to a stationary point.

\subsubsection*{3) Update of FIRES Element Positions}

With $\{\mathbf w_{\mathrm R,k}\}_{k=1}^{K}$, $\{\mathbf w_{\mathrm T,q}\}_{q=1}^{Q}$,
$\boldsymbol{\Phi}_{\mathrm b}$, and $\boldsymbol{\Theta}_{\mathrm b}$ fixed,
the FIRES element positions $\mathbf p$ are optimized by Algorithm~1,
which combines the Dinkelbach updates \eqref{34}, \eqref{56} and the
MM framework in
\eqref{41}--\eqref{43} and \eqref{Q_upp}--\eqref{P3C},
with the elements updated sequentially.

For a given feasible point $\mathbf p_l^{(t)}$, Lemma~\ref{lemma:taylor}
constructs a concave lower bound $\tilde{\mathcal C}(\mathbf p_l\mid \mathbf p_l^{(t)})$
and a convex upper bound $\tilde{\mathcal D}(\mathbf p_l\mid \mathbf p_l^{(t)})$,
which are tight at $\mathbf p_l^{(t)}$, i.e.,
\begin{equation} \small
\label{eq:mm_pos}
\begin{aligned}
\tilde{\mathcal C}(\mathbf p_l\mid \mathbf p_l^{(t)}) &\le {\mathcal C}(\mathbf p_l), \quad
\tilde{\mathcal C}(\mathbf p_l^{(t)}\mid \mathbf p_l^{(t)}) = {\mathcal C}(\mathbf p_l^{(t)}),\\
\tilde{\mathcal D}(\mathbf p_l\mid \mathbf p_l^{(t)}) &\ge {\mathcal D}(\mathbf p_l), \quad
\tilde{\mathcal D}(\mathbf p_l^{(t)}\mid \mathbf p_l^{(t)}) = {\mathcal D}(\mathbf p_l^{(t)}).
\end{aligned}
\end{equation}
Hence, the reformulated constraints (e.g., \eqref{43} and \eqref{P3C})
are conservative convex approximations that preserve feasibility and are tight
at the current iterate \cite{7547360}.

With fixed auxiliary variables $\{y_{\mathrm R,k}^{(i)}\}$ and $\{y_{\mathrm T,q}^{(i)}\}$,
problem~\eqref{P3_OF1} is convex with respect to
$\mathbf p_l$, $\{\Xi_k\}$, and $\{\Upsilon_q\}$.
Let
\begin{equation} \small
J^{(i)} \triangleq
\sum_{k=1}^{K}\log_2(1+\Xi_k^{(i)})+
\sum_{q=1}^{Q}\log_2(1+\Upsilon_q^{(i)}).
\end{equation}
Since \eqref{P3_OF1} is solved to optimality for the current
$\{y_{\mathrm R,k}^{(i)}\}$ and $\{y_{\mathrm T,q}^{(i)}\}$, the inner loop yields
\begin{equation} \small
J^{(i+1)} \ge J^{(i)}.
\end{equation}
The updates \eqref{34} and \eqref{56} follow the Dinkelbach principle \cite{8314727} and converge
under the stopping rule in Algorithm~1.

Since each FIRES element is updated sequentially while keeping the others fixed,
and each update does not decrease the objective value,
the overall objective associated with $\mathbf p$ is non-decreasing.
Moreover, $\mathbf p$ is constrained within a bounded region and satisfies
the minimum-distance constraint.
Therefore, Algorithm~1 generates a bounded non-decreasing objective sequence
and is guaranteed to converge to a stationary point.

\subsubsection*{4) Overall Convergence}

Therefore, we obtain
\begin{equation} \small
\label{eq:mono_all}
R^{(t+1)} \ge R^{(t)}, \quad \forall t.
\end{equation}
Moreover, the objective function is upper-bounded due to the finite transmit
power budget and the presence of noise terms.
Therefore, the sequence $\{R^{(t)}\}$ is monotonically non-decreasing and bounded,
which guarantees the convergence of Algorithm~2.

\endgroup
\subsection{Complexity analysis}
The proposed algorithm involves three main subproblems:

\begin{itemize}
  \item \textbf{Beamforming Optimization}: Solved via SDP with \( (K+Q+3) \) semidefinite constraints and a cone of size \( M \times M \), the computational complexity is given by
$\mathcal{O}\left((K+Q+3) M^7 \log(1/\varepsilon_1)\right)$~\cite{5447068}.

  \item \textbf{FIRES Coefficient Optimization}: Based on two complex vectors of length $L$ and SCA, with per-iteration complexity $\mathcal{O}\left(L^3 (K+Q)^2 \log(1/\varepsilon_2)\right)$.
  
  \item \textbf{FIRES Position Optimization}: In Algorithm~1, each element is updated sequentially using MM, involving $K+Q$ users and $L$ elements, resulting in complexity $\mathcal{O}\left(K Q L^2 \log(1/\varepsilon_3)\right)$ \cite{grant2009cvx}.
\end{itemize}

Therefore,  the overall computational complexity of Algorithm~2 is
$\mathcal{O}((K+Q+3) M^7\log(1/\varepsilon_1)+L^3 (K+Q)^2 \log(1/\varepsilon_2)+(K Q L^2 \log(1/\varepsilon_3))$.

\begin{algorithm}[t]\small
\caption{Proposed Alternating Optimization Algorithm}
\begin{algorithmic}[1]
\STATE Initialize $n = 0$, $\{\mathbf{\Phi}_b^{(0)}, \mathbf{\Theta}_b^{(0)}\}$, and tolerance $\delta$.
\REPEAT
    \STATE \textbf{1) Beamforming Optimization:} 
    Solve problem \eqref{P1_OF3} via iterative updates until convergence.
    
    \STATE \textbf{2) FIRES Coefficients Optimization:} 
    Solve problem \eqref{P2_OF2} with fixed beamformers until convergence.

    \STATE \textbf{3) Element Position Optimization:} 
    Optimize $\mathbf{p}$ using Algorithm~1.

    \STATE $n \leftarrow n + 1$
\UNTIL the convergence criterion on the objective is satisfied.
\end{algorithmic}
\end{algorithm}

\section{Simulation Results}
{
In this simulation, the proposed FIRES-NOMA system operating at a carrier frequency of $10$ GHz, corresponding to a wavelength of $\lambda = 0.03$ m. The BS employs a ULA with $M = 8$ antennas and serves four single-antenna users, including $K = 2$ users located in the reflection region and $Q = 2$ users in the transmission region. The FIRES is composed of $N = 25$ elements that are initially arranged uniformly within a square aperture of side length $4.5\lambda$ to ensure sufficient space for element movement, with a minimum inter-element spacing of $\Delta_D = \lambda/2$. The azimuth angle from the BS to the FIRES is set to $\phi_{\text{t}} = 120^\circ$, while the azimuth and elevation angles at the FIRES for receiving the signal from the BS are set to $\phi_{\text{r}} = 330^\circ$ and $\psi_{\text{r}} = 30^\circ$, respectively. The FIRES transmits signals toward the transmission users with azimuth angles $\phi_{\text{T},q} = [140^\circ, 210^\circ]$, and toward the reflection users with $\phi_{\text{R},k} = [-45^\circ, 30^\circ]$. All users are assigned an elevation angle of $\psi_{\text{T},q} = \psi_{\text{R},k}=-30^\circ$. The direct path azimuth angles from the BS to users are configured as $[100^\circ, 130^\circ]$. The distance between the BS and the FIRES is 70 m, while the FIRES-to-user distances are 5 m and 3 m for the transmission users, and 15 m and 30 m for the reflection users, respectively. All channels follow a Rician fading model with $\kappa$ = 1 \cite{wu2023two}. In addition, the channel fading coefficients $\zeta_{\mathrm{G}}$, $\zeta_{\mathrm{b},k}$, $\zeta_{\mathrm{R},k}$ and $\zeta_{\mathrm{T},q}$ are calculated by $\rho_0  \left( \frac{1}{\mathrm{dist}} \right)^2
$, where $\rho_0=-30$ dB denotes the reference path loss coefficient, and $\mathrm{dist}$ represents the distance between the transmitter and the receiver \cite{STAR-2}. The noise power is set to $-80$ dBm, and the maximum transmit power of the BS is $P_{\text{max}}$ = 10 dBm. Each user is required to meet a minimum QoS threshold of $\gamma_{\text{min}}$ = 1 bps/Hz. }

To validate the effectiveness of the proposed approach, the following benchmark schemes are considered for comparison. In what follows, we denote the proposed FIRES-aided optimization framework as \textbf{FIRES}, which will be consistently used as its label in all subsequent figures and discussions.

\begin{itemize}
\item \textbf{STAR:} {Conventional STAR-RIS elements are uniformly distributed over a fixed surface without position optimization, which represents a special case of the proposed framework when the FIRES element positions are fixed. All other settings, including the BS beamforming and the transmission/reflection coefficient design, are identical to those in the proposed method.}

\item \textbf{D-FIRES:} FIRES elements are selected from a discrete candidate set with spacing satisfying the minimum distance constraint. A greedy search is used to sequentially place elements for maximum objective gain. Other configurations remain unchanged.

\item \textbf{STAR (OMA):} Conventional STAR-RIS operating with OMA instead of NOMA, while all other procedures strictly follow those of the proposed algorithm.

\end{itemize}

\begin{figure}[t]
    \centering
    \includegraphics[width=0.4\textwidth]{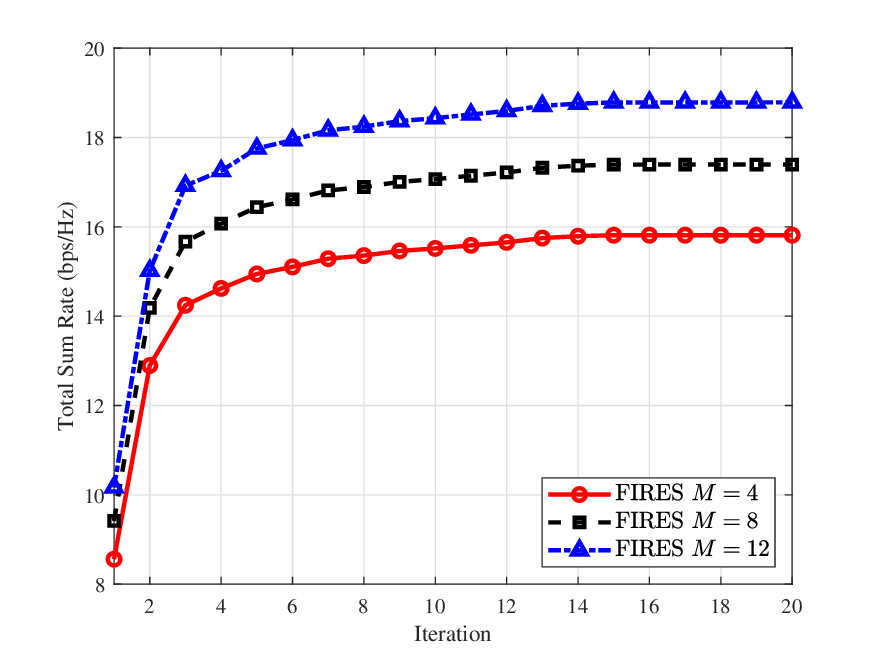}
    \caption{Validation of Convergence for Algorithm 2.}
    \label{fig:example}
\end{figure}

As shown in Fig.~2, the proposed algorithm achieves convergence within approximately 12 iterations, indicating its desirable convergence efficiency. In addition, the three curves correspond to different numbers of BS antennas ($M = 4, 8, 12$). It is observed that increasing the number of antennas results in a higher total sum rate, validating the benefit of additional antenna resources in improving system performance.

\begin{figure}[t]
    \centering
    \includegraphics[width=0.4\textwidth]{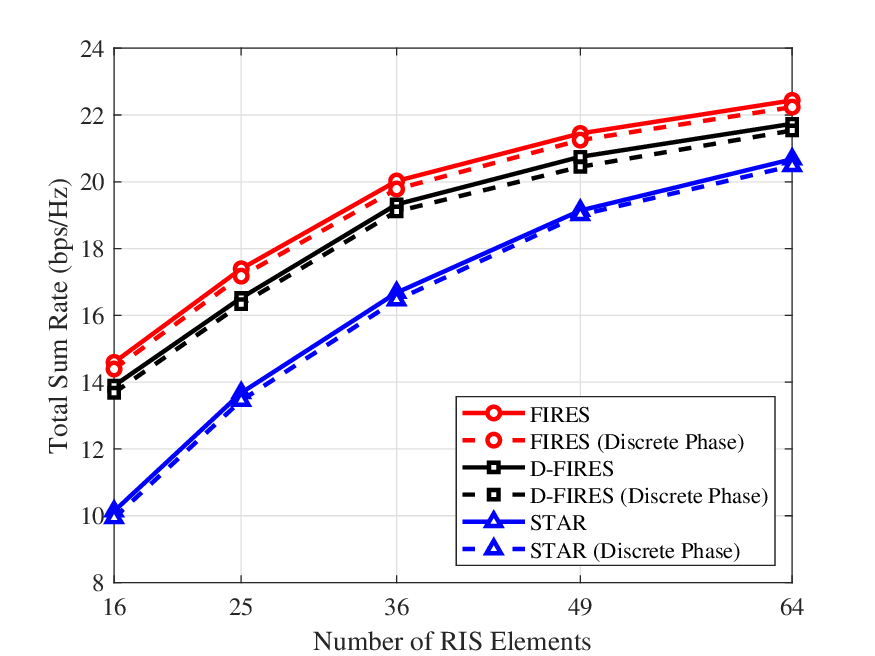}
    \caption{{Total sum rate comparison with the different number of RIS elements.}}
    \label{fig:example}
            \vspace{-1em}
\end{figure}

Fig.~3 illustrates the impact of the number of RIS elements \( L \) on the total sum rate performance. It can be observed that the proposed scheme and all benchmark schemes exhibit improved performance as \( L \) increases. Notably, the proposed method consistently outperforms the benchmarks across all considered values of \( L \), demonstrating the effectiveness of jointly optimizing the FIRES element positions.
Moreover, it is worth highlighting that the proposed scheme achieves a performance comparable to the conventional STAR-RIS with \( L = 64 \) using only \( L = 36 \) elements. This significant performance gain underscores the importance of position optimization, which enables the system to achieve higher efficiency with fewer hardware resources, thereby reducing deployment costs.
Additionally, the proposed continuous FIRES outperforms the discrete-position benchmark (D-FIRES) by approximately 0.8~dB in the considered scenario, confirming that continuous element positioning yields notable performance gains over predefined discrete placements. From a practical implementation perspective, Fig.~3 also includes the performance of a discrete-phase realization, where the FIRES phase shifts are quantized using a 3-bit resolution. The results show that the performance loss introduced by phase discretization is marginal, with an approximately 
0.2~dB degradation compared to the continuous-phase case. This observation indicates that the proposed FIRES-assisted design does not critically rely on ideal continuous phase control and remains effective under finite-resolution phase shifters, further supporting its practical feasibility. {In addition, optimizing beamforming or FIRES coefficients alone yields limited performance, demonstrating that the joint optimization of beamforming, FIRES coefficients, and position is essential to fully unlock the potential of the proposed system.}

\begin{figure}[t]
    \centering
    \includegraphics[width=0.4\textwidth]{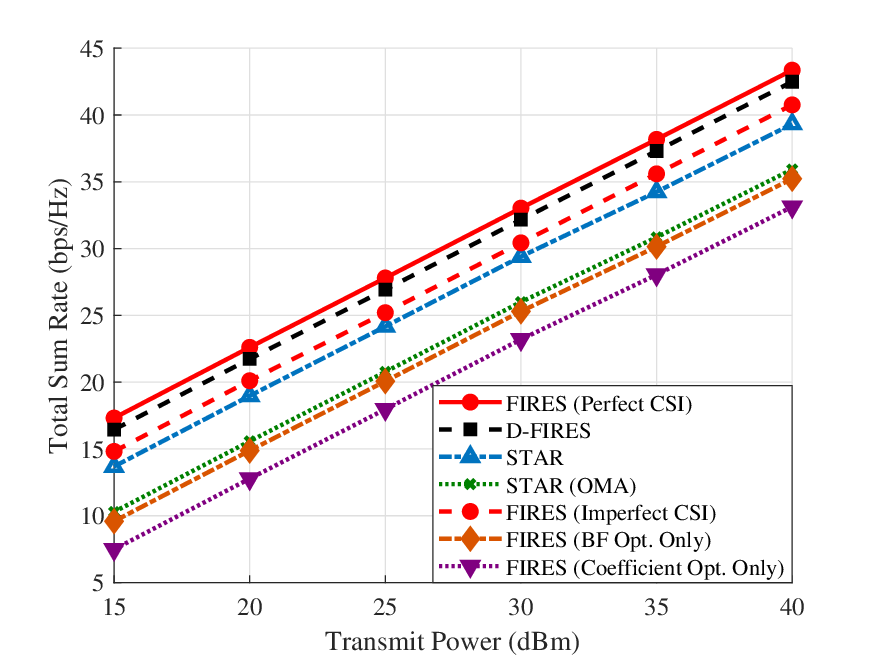}
    \caption{{Total sum rate comparison with different transmit power at the BS.}}
    \label{fig:example}
            \vspace{-1em}
\end{figure}
Fig.~4 illustrates the total sum rate performance of different schemes under varying BS transmit power levels. As expected, the proposed FIRES scheme consistently outperforms all benchmarks. Specifically, when the transmit power is set to 15~dBm, the proposed method achieves a 27\% improvement in system throughput compared to the conventional STAR-RIS scheme, demonstrating the effectiveness of FIRES element position optimization. Moreover, to achieve the same performance as the conventional STAR-RIS at 20~dBm, the proposed method only requires approximately 15~dBm transmit power, indicating its potential for energy savings. Additionally, we compare the proposed system with the conventional STAR-RIS-assisted OMA scheme. It is evident that OMA exhibits significantly lower performance compared to all three NOMA-based schemes, due to its inefficient spectrum usage and severe multi-user interference. { To evaluate the robustness of the proposed design, we consider imperfect CSI
for all channel links involved in the joint optimization and SIC decoding,
including the BS--FIRES channel $\mathbf{G}$, the direct BS--user channel
$\mathbf{h}_{\text{b},k}$, and the FIRES--user channels
$\mathbf{h}_{\mathrm{R},k}$ and $\mathbf{h}_{\mathrm{T},q}$. Specifically, for a
generic channel matrix/vector
$\mathbf{H}\in\{\mathbf{G}, \mathbf{h}_{\text{b},k},
\mathbf{h}_{\mathrm{R},k}, \mathbf{h}_{\mathrm{T},q}\}$,
its imperfect estimate is modeled as
$
\hat{\mathbf{H}}
=
\sqrt{1-\varepsilon}\,\mathbf{H}
+
\sqrt{\varepsilon}\,\tilde{\mathbf{H}},
$
where $\varepsilon\in[0,1]$ denotes the channel estimation error variance, and
$\tilde{\mathbf{H}}$ represents the channel estimation error. The entries of
$\tilde{\mathbf{H}}$ are modeled as independent complex Gaussian random variables
following $[\tilde{\mathbf{H}}]_{\ell}\sim\mathcal{CN}\!\left(0,\,\big|[\mathbf{H}]_{\ell}\big|^2\right), \quad \forall \ell,$
which corresponds to a relative CSI error model where the error
power scales with the channel strength. This statistical model is widely
adopted in the literature to evaluate the robustness of RIS-assisted systems
under imperfect CSI. Unless otherwise specified, we set
$\varepsilon=0.1$ throughout the simulations \cite{9732214}. {The results indicate that imperfect CSI leads to a performance loss compared with the perfect-CSI benchmark at the considered error level. Nevertheless, the proposed system continues to deliver clear performance
improvements under imperfect CSI, which confirms its effectiveness in
practical scenarios.}

\begin{figure}[t]
    \centering
    \includegraphics[width=0.4\textwidth]{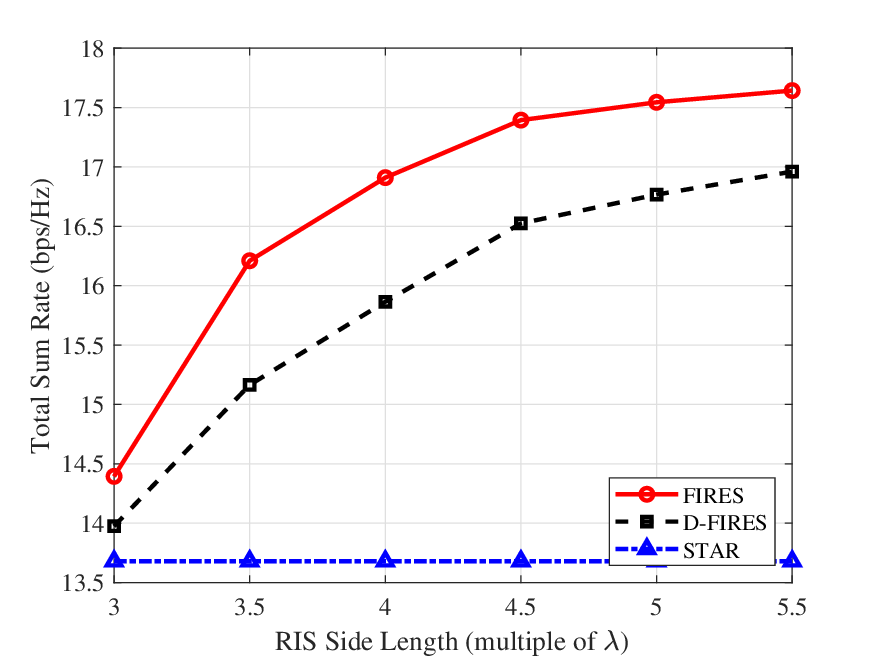}
    \caption{Total sum rate comparison with different RIS side length.}
    \label{fig:example}
            \vspace{-1em}
\end{figure}

Fig.~5 presents the total sum rate comparison under varying RIS side lengths. To facilitate understanding, we first clarify the underlying principle: in the conventional STAR-RIS scheme, as the side length increases, the element positions remain unchanged, maintaining the predefined half-wavelength spacing. In contrast, the proposed FIRES scheme and D-FIRES scheme dynamically optimize the element positions based on the available aperture area. As shown in the figure, increasing the side length provides greater spatial degrees of freedom, enabling further performance gains. The proposed scheme consistently outperforms the discrete-position-based D-FIRES across all side lengths, with the most notable gain of approximately 1.5~dB observed at a side length of $3.5\lambda$. This highlights the advantage of continuous position optimization in fully exploiting the RIS surface. Moreover, beyond $4.5\lambda$, the rate improvement becomes marginal while the implementation and computational costs grow significantly, suggesting that practical designs should consider the trade-off between performance and cost.

\begin{figure}[t]
    \centering
    \includegraphics[width=0.4\textwidth]{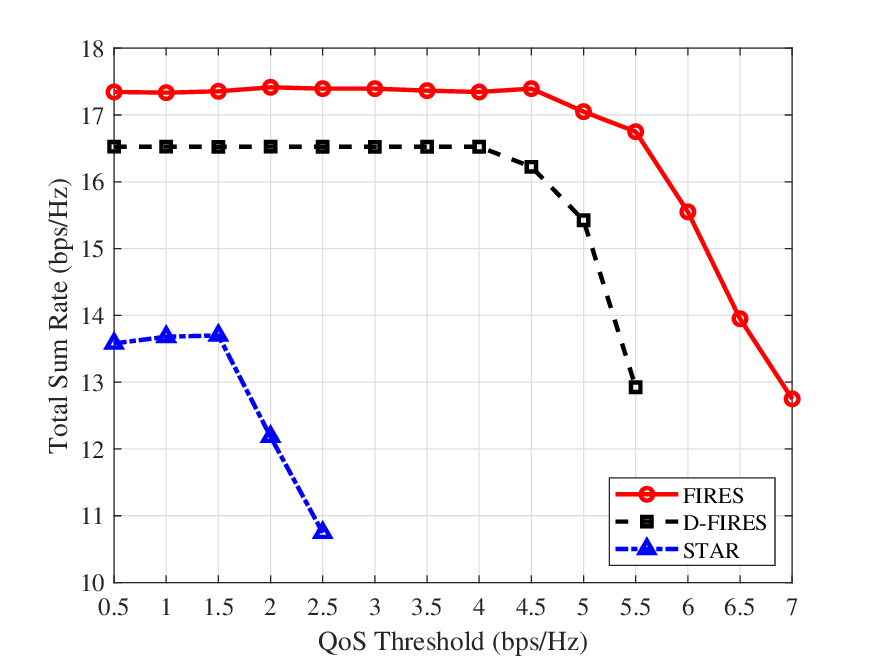}
    \caption{Total sum rate comparison with different Qos threshold}
    \label{fig:example}
            \vspace{-1em}
\end{figure}

Fig.~6 depicts the total sum rate performance under different QoS thresholds per user. It can be observed that the proposed FIRES scheme consistently achieves the highest performance compared with all baseline schemes. As the minimum QoS requirement increases, the total sum rate initially remains stable. However, once the QoS threshold exceeds a certain level, the system begins to experience a sharp performance drop due to the increasing difficulty of satisfying all user constraints. Eventually, each scheme reaches a point beyond which no feasible solution exists, and the algorithm ceases to operate.
Notably, the proposed FIRES scheme can accommodate user-specific QoS requirements up to 7~bps/Hz, whereas STAR and D-FIRES can only support up to 2.5 and 5.5~bps/Hz, respectively. This significant advantage arises from the additional spatial DoF introduced by the continuous position optimization in FIRES, which enables more flexible beamforming and improved interference mitigation. As a result, FIRES demonstrates superior capability in maintaining service quality under stringent QoS demands.

\section{Conclusion}
This paper proposes a FIRES-assisted NOMA multi-user communication system. In the proposed system, FIRES elements support simultaneous transmission and reflection and are able to move flexibly within a predefined area, thereby enhancing spatial control and adaptability. To maximize the total sum rate, a non-convex optimization problem is formulated and solved via a three-layer alternating optimization algorithm, which incorporates an MM framework to handle position optimization under complex channel coupling. Simulation results demonstrate that the proposed system achieves up to a 27\% higher sum rate compared to conventional STAR-RIS systems with the same number of elements, or equivalently achieves the same performance with nearly 50\% fewer elements, thereby highlighting its potential for cost-efficient large-scale deployment. Furthermore, the findings also confirm that continuous element positioning provides higher performance than discrete configurations, underscoring the benefit of exploiting continuous spatial freedom in FIRES. Future research directions include more refined electromagnetic-level modeling to capture mutual coupling and scattering effects associated with element mobility, as well as more advanced joint or block-wise position optimization strategies to improve efficiency.}


\appendices
\section{Derivation of \eqref{A-a}} \label{appendix:derivation}
For the term $|\mathbf{F}_{\text{R},k} \mathbf{w}_\mu|^2$, it can be expanded in a manner similar to the transformation presented in \eqref{23}, as follows:

\begin{equation} \small
\begin{aligned}
&|\mathbf{F}_{\text{R},k} \mathbf{w}_\mu|^2=  {\mathbf{h}}_{\text{R},k}^H {\mathbf{\Phi}}_{\text{b}}^H {\mathbf{G}}\mathbf{w}_\mu \mathbf{w}_\mu^H
{\mathbf{G}}^H
{\mathbf{\Phi}}_{\text{b}} {\mathbf{h}}_{\text{R},k}+\mathbf{h}_{\text{b},k}\mathbf{w}_\mu \mathbf{w}_\mu^H
{\mathbf{G}}^H\\&{\mathbf{\Phi}}_{\text{b}} {\mathbf{h}}_{\text{R},k} 
+
{\mathbf{h}}_{\text{R},k}^H {\mathbf{\Phi}}_{\text{b}}^H {\mathbf{G}}\mathbf{w}_\mu \mathbf{w}_\mu^H\mathbf{h}_{\text{b},k}^H
+\mathbf{h}_{\text{b},k}\mathbf{w}_\mu \mathbf{w}_\mu^H\mathbf{h}_{\text{b},k}^H
\label{35}.
\end{aligned}
\end{equation}

Similarly, since all four extended terms in \eqref{35} are scalars and hence equal to their respective traces, the expression can be equivalently rewritten using the cyclic property of the trace as follows:
\vspace{0.02em}
\begin{equation} \small
\begin{aligned}
|\mathbf{F}_{\text{R},k} \mathbf{w}_\mu|^2= \mathbf{A}_{\mu,\text{1}}+\mathbf{A}_{\mu,\text{2}}+\mathbf{A}_{\mu,\text{3}}+\mathbf{A}_{\mu,\text{4}},\label{wod}
\end{aligned}
\end{equation}
where $\mathbf{A}_{\mu,\text{1}} =  \mathbf{w}_\mu^H
{\mathbf{G}}^H
{\mathbf{\Phi}}_{\text{b}} {\mathbf{h}}_{\text{R},k} {\mathbf{h}}_{\text{R},k}^H {\mathbf{\Phi}}_{\text{b}}^H {\mathbf{G}}\mathbf{w}_\mu $, $\mathbf{A}_{\mu,\text{2}} =  \mathbf{w}_\mu^H
{\mathbf{G}}^H{\mathbf{\Phi}}_{\text{b}} {\mathbf{h}}_{\text{R},k}\mathbf{h}_{\text{b},k}\mathbf{w}_\mu $,
$\mathbf{A}_{\mu,\text{3}}=\mathbf{w}_\mu^H\mathbf{h}_{\text{b},k}^H{\mathbf{h}}_{\text{R},k}^H {\mathbf{\Phi}}_{\text{b}}^H {\mathbf{G}}\mathbf{w}_\mu $ and $\mathbf{A}_{\mu,\text{4}} =  \mathbf{w}_\mu^H\mathbf{h}_{\text{b},k}^H\mathbf{h}_{\text{b},k}\mathbf{w}_\mu$.

The outer function of $\mathbf{A}_{\mu,\text{1}}$ is processed by substituting \eqref{7} and\eqref{9} into the expression of $\mathbf{A}_{\mu,\text{1}}$, followed by removing terms that are not related to
$\mathbf p$. We can rewrite the expression of $\mathbf{A}_{\mu,\text{1}}$ as
\begin{equation} \small
\begin{aligned}
\mathbf{A}_{\mu,\text{1}}
&=Z_{\mu,\text{1}}\mathbf{a}_{\text{r}}(\mathbf{p})^H
{\mathbf{\Phi}}_{\text{b}} \mathbf{a}_{\text{R},k}(\mathbf{p})  \mathbf{a}_{\text{R},k}^H(\mathbf{p}) {\mathbf{\Phi}}_{\text{b}}^H \mathbf{a}_{\text{r}}(\mathbf{p}) \\
&=Z_{\mu,\text{1}}\mathbf{a}_{\text{r}}(\mathbf{p})^H
\overline{\mathbf{a}}_{\text{R},k}(\mathbf{p})\mathbf{v}_2\mathbf{v}_2^H\overline{\mathbf{a}}_{\text{R},k}(\mathbf{p})^H\mathbf{a}_{\text{r}}(\mathbf{p}) ,\label{36}
\end{aligned}
\end{equation}
where $Z_{\mu,\text{1}}=\mathbf{a}_{\text{t}}^H\mathbf{w}_\mu{\zeta_{\text{G}}} {\zeta_{\text{R},k} }\mathbf{w}_\mu^H
 \mathbf{a}_{\text{t}}$, and $\overline{\mathbf{a}}_{\text{R},k}(\mathbf{p})=\text{diag}(\mathbf{a}_{\text{R},k}(\mathbf{p}) )$.

Next, we turn our attention to \( \mathbf{A}_{\mu,2} \) and \( \mathbf{A}_{\mu,3} \). Similar to the approach in \eqref{36}, the expressions of \( \mathbf{A}_{\mu,2} \) and \( \mathbf{A}_{\mu,3} \) can be rewritten as follows:
\begin{equation} \small
\begin{aligned}
\mathbf{A}_{\mu,\text{2}}& =  \mathbf{w}_\mu^H
\sqrt{\zeta_{\text{G}}} \mathbf{a}_{\text{t}}\mathbf{a}_{\text{r}}(\mathbf{p})^H{\mathbf{\Phi}}_{\text{b}}   \mathbf{a}_{\text{R},k}( \mathbf{p})\sqrt{\zeta_{\text{R},k} } \mathbf{h}_{\text{b},k}\mathbf{w}_\mu \\&
=
Z_{\mu,\text{2}}\overline{\mathbf{a}}_{r}(\mathbf{p})\mathbf{a}_{\text{R},k}( \mathbf{p}),\label{41A}
\end{aligned}
\end{equation}
\begin{equation} \small
\begin{aligned}
\mathbf{A}_{\mu,\text{3}}&=\mathbf{w}_\mu^H\mathbf{h}_{\text{b},k}^H\sqrt{\zeta_{\text{R},k} } \mathbf{a}_{\text{R},k}^H( \mathbf{p}){\mathbf{\Phi}}_{\text{b}}^H \sqrt{\zeta_{\text{G}}} \mathbf{a}_{\text{r}}(\mathbf{p}) \mathbf{a}_{\text{t}}^H\mathbf{w}_\mu 
\\&
=Z_{\mu,\text{3}}\mathbf{a}_{\text{R},k}^H( \mathbf{p})\overline{\mathbf{a}}_{r}^H(\mathbf{p}),\label{42}
\end{aligned}
\end{equation}
where \( Z_{\mu,2} = \sqrt{\zeta_{\text{R},k}} \sqrt{\zeta_{\text{G}}} \, \mathbf{h}_{\text{b},k} \mathbf{w}_{\mu} \mathbf{w}_{\mu}^H \mathbf{a}_{\text{t}} \mathbf{v}_2^T \), and  
\( Z_{\mu,3} = \sqrt{\zeta_{\text{R},k}} \sqrt{\zeta_{\text{G}}} \, \mathbf{v}_2^* \mathbf{a}_{\text{t}}^H \mathbf{w}_{\mu} \mathbf{w}_{\mu}^H \mathbf{h}_{\text{b},k}^H \).  
Note that since \( \mathbf{w}_{\mu} \mathbf{w}_{\mu}^H \) is a Hermitian matrix and the remaining terms are conjugate transposes of each other, we have  
$Z_{\mu,2} = Z_{\mu,3}^H$
.
Additionally, \( \overline{\mathbf{a}}_{r}(\mathbf{p}) \) is defined as  
$
\overline{\mathbf{a}}_{r}(\mathbf{p}) = \operatorname{diag}(\mathbf{a}_{r}^H(\mathbf{p})).
$

Therefore, according to the expression above, we can rewrite \eqref{wod} as
\begin{equation} \small
\begin{aligned}
|\mathbf{F}_{\text{R},k} \mathbf{w}_\mu|^2&= \mathbf{A}_{\mu,\text{1}}+\mathbf{A}_{\mu,\text{2}}+\mathbf{A}_{\mu,\text{3}}+\mathbf{A}_{\mu,\text{4}}\\&
=Z_{\mu,\text{1}}\mathbf{a}_{\text{r}}(\mathbf{p})^H
\overline{\mathbf{a}}_{\text{R},k}(\mathbf{p})\mathbf{v}_2\mathbf{v}_2^H\overline{\mathbf{a}}_{\text{R},k}(\mathbf{p})^H\mathbf{a}_{\text{r}}(\mathbf{p})\\+&Z_{\mu,\text{2}}\overline{\mathbf{a}}_{r}(\mathbf{p})\mathbf{a}_{\text{R},k}( \mathbf{p})+Z_{\mu,\text{3}}\mathbf{a}_{\text{R},k}^H( \mathbf{p})\overline{\mathbf{a}}_{r}^H(\mathbf{p})+\mathbf{A}_{\mu,\text{4}}.
\end{aligned}
\end{equation}

This completes the derivation.

\section{Derivation of \eqref{ck2} and \eqref{dk2}} \label{appendix:derivation2}
The detailed derivation is given based on $|\mathbf{F}_{\text{R},k} \mathbf{w}_\mu|^2$. To begin with, $\mathbf{A}_{\mu,\text{1}}$ can be rewritten as
\vspace{-1.0ex}
\begin{equation} \small
\begin{aligned}
\mathbf{A}_{\mu,\text{1}} = &Z_{\mu,\text{1}}\sum_{i=1}^{L} \sum_{j=1}^{L}
\left[ \mathbf{v}_2\mathbf{v}_2^H\right]_{i,j} \cdot 
\left[ \mathbf{a}_r^H(\mathbf{p}) \, \overline{\mathbf{a}}_{\text{R},k}(\mathbf{p})\right]_i \cdot
\left[ \overline{\mathbf{a}}_{\text{R},k}^H(\mathbf{p}) \, \mathbf{a}_r(\mathbf{p}) \right]_j\\
=&Z_{\mu,\text{1}}\sum_{i=1}^{L} \sum_{j=1}^{L}
\left[ \mathbf{v}_2\mathbf{v}_2^H\right]_{i,j} \\&
\cdot e^{-j \frac{2\pi}{\lambda} \left( d_{r,i}(\mathbf{p}_i) - d_{\text{S},k,i}(\mathbf{p}_i) \right)} \cdot 
e^{j \frac{2\pi}{\lambda} \left( d_{r,j}(\mathbf{p}_j) - d_{\text{S},k,j}(\mathbf{p}_j) \right)},\label{66}
\end{aligned}
\end{equation}
where $d_{r,l}(\mathbf{p}_l)  = p_{x,l} \sin(\phi_{r}) \cos(\psi_r) + p_{y,l} \sin(\psi_{r})$ and $d_{\text{S},k,l}(\mathbf{p}_l) = p_{x,l} \sin(\phi_{\text{R},k}) \cos(\psi_{\text{R},k}) + p_{y,l} \sin(\psi_{\text{R},k}).$
It is evident that when \( i = j \), the complex exponential components cancel each other out, resulting in constant terms that are independent of the position variable \( \mathbf{p} \). Moreover, since \(  \mathbf{v}_2\mathbf{v}_2^H\) is a Hermitian matrix, the terms with \( i \neq j \) appear in conjugate pairs. Leveraging this structure, and by focusing on the \( l \)th element (i.e., when \( i = l \) or \( j = l \)) while discarding the constant terms associated with \( i = j \),  the optimization-equivalent form of \eqref{66} is given by
\vspace{-1.0ex}
\begin{equation} \small
\begin{aligned}
\mathbf{A}_{\mu,\text{1}}(\mathbf{p}_l)& \equiv  Z_{\mu,\text{1}} \sum_{\mathclap{i=1,\; i \neq l}}^{L}
2\Re \{ \left[\mathbf{v}_2\mathbf{v}_2^H \right]_{i,l}\\& \cdot 
e^{-j \frac{2\pi}{\lambda} \left( d_{r,i}(\mathbf{p}_i) - d_{\text{S},k,i}(\mathbf{p}_i) \right)} \cdot 
e^{j \frac{2\pi}{\lambda} \left( d_{r,l}(\mathbf{p}_l) - d_{\text{S},k,l}(\mathbf{p}_l) \right)}\}.\label{38}
\end{aligned}
\end{equation}
It can be observed that the term 
\( \left[ \mathbf{v}_2 \mathbf{v}_2^H \right]_{i,l} \cdot 
e^{-j \frac{2\pi}{\lambda} \left( d_{r,i}(\mathbf{p}_i) - d_{\text{S},k,i}(\mathbf{p}_i) \right)} \)
is independent of \( \mathbf{p}_l \) when only the \( l \)th element is considered. Therefore, it can be treated as a constant, denoted by \( s_{i,l,k} \). Since the amplitude of \( s_{i,l,k} \) is equal to 1,  \eqref{66} can be further simplified as
\begin{equation} \small
\begin{aligned}
\mathbf{A}_{\mu,\text{1}}(\mathbf{p}_l)& \equiv Z_{\mu,\text{1}} \sum_{\mathclap{i=1,\; i \neq l}}^{L}
2\Re \{ s_{i,l,k}
e^{j \frac{2\pi}{\lambda} \left( d_{r,l}(\mathbf{p}_l) - d_{\text{S},k,l}(\mathbf{p}_l) \right)}\}
\\&\equiv Z_{\mu,\text{1}} \sum_{\mathclap{i=1,\; i \neq l}}^{L}\cos{(\angle s_{i,l,k} +\frac{2\pi}{\lambda}\Delta d_{k,l}(\mathbf{p}_l)) },\label{39}
\end{aligned}
\end{equation}
where $\angle s_{i,l,k}$ denotes the phase of $s_{i,l,k}$, and $\Delta d_{k,l}(\mathbf{p}_l)$ denotes the term of $\left( d_{r,l}(\mathbf{p}_l) - d_{\text{S},k,l}(\mathbf{p}_l) \right)$. 

Next, by optimizing the position of one FIRES element at a time, we further simplify \eqref{41A} and \eqref{42} as
\begin{equation} \small
\begin{aligned}
&\mathbf{A}_{\mu,\text{2}}(\mathbf{p}_l)=
\tilde Z_{\mu,\text{2},l}e^{-j \frac{2\pi}{\lambda} \left( d_{r,l}(\mathbf{p}_l) - d_{\text{S},k,l}(\mathbf{p}_l) \right)}
=\tilde Z_{\mu,\text{2},l}e^{-j \frac{2\pi}{\lambda} \Delta d_{k,l}(\mathbf{p}_l)}\\
=&\tilde Z_{\mu,\text{2},l}\left(\cos(-\frac{2\pi}{\lambda}\Delta d_{k,l}(\mathbf{p}_l))+j\sin(-\frac{2\pi}{\lambda}\Delta d_{k,l}(\mathbf{p}_l))\right),
\end{aligned}
\end{equation}
\begin{equation} \small
\begin{aligned}
&\mathbf{A}_{\mu,\text{3}}(\mathbf{p}_l) =
\tilde Z_{\mu,\text{3},l}e^{j \frac{2\pi}{\lambda} \left( d_{r,l}(\mathbf{p}_l) - d_{\text{S},k,l}(\mathbf{p}_l) \right)}
=\tilde Z_{\mu,\text{3},l}e^{j \frac{2\pi}{\lambda} \Delta d_{k,l}(\mathbf{p}_l)}\\&=\tilde Z_{\mu,\text{3},l}\left( \cos(\frac{2\pi}{\lambda}\Delta d_{k,l}(\mathbf{p}_l))+j\sin(\frac{2\pi}{\lambda}\Delta d_{k,l}(\mathbf{p}_l))\right) ,
\end{aligned}
\end{equation}
where $\tilde Z_{\mu,\text{2},l}=\left [Z_{\mu,\text{2}}\right ]_l$ and $\tilde Z_{\mu,\text{3},l}=\left [Z_{\mu,\text{3}}\right ]_l$. 

Define $\mathbf{T}_{\mu}(\mathbf{p}_l) =\mathbf{A}_{\mu,\text{2}}(\mathbf{p}_l)+\mathbf{A}_{\mu,\text{3}}(\mathbf{p}_l) $, we further simplify the expression by rewriting $\mathbf{T}_{\mu}$. Since $\tilde Z_{\mu,\text{2},l}=\tilde Z_{\mu,\text{3},l}^H$ and both are scalar values, they can be expressed as  
 $\tilde Z_{\mu,\text{2},l} = \Re(\tilde Z_{\mu,\text{2},l})+j\Im(\tilde Z_{\mu,\text{2},l})$, and $\tilde Z_{\mu,\text{3},l} = \Re(\tilde Z_{\mu,\text{2},l})-j\Im(\tilde Z_{\mu,\text{2},l})$. Applying the standard trigonometric properties $\cos(x)=\cos(-x)$
and $\sin(-x)=-\sin(x)$, we obtained:
\begin{equation} \small
\begin{aligned}
&\mathbf{T}_{\mu}(\mathbf{p}_l)=\left(\Re(\tilde Z_{\mu,\text{2},l})+j\Im(\tilde Z_{\mu,\text{2},l})
\right)\cos(\frac{2\pi}{\lambda}\Delta d_{k,l}(\mathbf{p}_l))
\\&
-\left(\Re(\tilde Z_{\mu,\text{2},l})+j\Im(\tilde Z_{\mu,\text{2},l})
\right)j\sin(\frac{2\pi}{\lambda}\Delta d_{k,l}(\mathbf{p}_l))
\\&+
\left(\Re(\tilde Z_{\mu,\text{2},l})-j\Im(\tilde Z_{\mu,\text{2},l})\right) \cos(\frac{2\pi}{\lambda}\Delta d_{k,l}(\mathbf{p}_l))\\&
+\left(\Re(\tilde Z_{\mu,\text{2},l})-j\Im(\tilde Z_{\mu,\text{2},l})\right)j\sin(\frac{2\pi}{\lambda}\Delta d_{k,l}(\mathbf{p}_l))\\=&
2\Re(\tilde Z_{\mu,\text{2},l}) \cos(\frac{2\pi}{\lambda}\Delta d_{k,l}(\mathbf{p}_l))
+2\Im(\tilde Z_{\mu,\text{2},l})\sin(\frac{2\pi}{\lambda}\Delta d_{k,l}(\mathbf{p}_l)).\label{45}
\end{aligned}
\end{equation}

This completes the derivation.

\section{Derivation of the Gradient and Hessian in \eqref{41}} \label{appendix:derivation3}

This appendix provides the detailed derivations of the gradient and Hessian matrices of $\mathbf{A}_{\mu,\text{1}}(\mathbf{p}_l)$ and $\mathbf{T}_{\mu}(\mathbf{p}_l)$ with respect to the position of the $l$th FIRES element, denoted by $\mathbf{p}_l = [p_{x,l}, p_{y,l}]^T$.

To simplify the notation, we define the directional index $d \in \{x, y\}$ and the corresponding coefficients as
$\hat{D}_{x,k} = \frac{2\pi}{\lambda} \left( \sin(\phi_{r}) \cos(\psi_r) - \sin(\phi_{\text{R},k}) \cos(\psi_{\text{R},k}) \right), \quad 
\hat{D}_{y,k} = \frac{2\pi}{\lambda} \left( \sin(\phi_{r}) - \sin(\phi_{\text{R},k}) \right).$

Then, the partial derivatives of $\mathbf{A}_{\mu,\text{1}}(\mathbf{p}_l)$ are expressed as{\small
\begin{subequations} 
\begin{align}
\frac{\partial \mathbf{A}_{\mu,\text{1}}(\mathbf{p}_l)}{\partial p_{d,l}} &= -Z_{\mu,\text{1}} \hat{D}_{d,k} \sum_{\mathclap{i=1,\; i \neq l}}^{L} \sin\left( \angle s_{i,l,k} + \frac{2\pi}{\lambda} \Delta d_{k,l}(\mathbf{p}_l) \right), \\
\frac{\partial^2 \mathbf{A}_{\mu,\text{1}}(\mathbf{p}_l)}{\partial p_{d,l}^2} &= -Z_{\mu,\text{1}} \hat{D}_{d,k}^2 \sum_{\mathclap{i=1,\; i \neq l}}^{L} \cos\left( \angle s_{i,l,k} + \frac{2\pi}{\lambda} \Delta d_{k,l}(\mathbf{p}_l) \right), \\
\frac{\partial^2 \mathbf{A}_{\mu,\text{1}}(\mathbf{p}_l)}{\partial p_{x,l} \partial p_{y,l}} &= -Z_{\mu,\text{1}} \hat{D}_{x,k} \hat{D}_{y,k} \sum_{\mathclap{i=1,\; i \neq l}}^{L} \cos\left( \angle s_{i,l,k} + \frac{2\pi}{\lambda} \Delta d_{k,l}(\mathbf{p}_l) \right).
\end{align}
\end{subequations}}

The gradient and Hessian matrix are then assembled as
\[
\begin{aligned}
\nabla \mathbf{A}_{\mu,\text{1}}(\mathbf{p}_l) &= 
\begin{bmatrix}
\frac{\partial \mathbf{A}_{\mu,\text{1}}}{\partial p_{x,l}} &
\frac{\partial \mathbf{A}_{\mu,\text{1}}}{\partial p_{y,l}}
\end{bmatrix}^T, \\
\nabla^2 \mathbf{A}_{\mu,\text{1}}(\mathbf{p}_l) &= 
\begin{bmatrix}
\frac{\partial^2 \mathbf{A}_{\mu,\text{1}}}{\partial p_{x,l}^2} & \frac{\partial^2 \mathbf{A}_{\mu,\text{1}}}{\partial p_{x,l} \partial p_{y,l}} \\
\frac{\partial^2 \mathbf{A}_{\mu,\text{1}}}{\partial p_{y,l} \partial p_{x,l}} & \frac{\partial^2 \mathbf{A}_{\mu,\text{1}}}{\partial p_{y,l}^2}
\end{bmatrix}.
\end{aligned}
\]

Similarly, the partial derivatives of $\mathbf{T}_{\mu}(\mathbf{p}_l)$ are given by
\begin{subequations} \small
\begin{align}
\frac{\partial \mathbf{T}_{\mu}(\mathbf{p}_l)}{\partial p_{d,l}} &= 
-2\Re(\tilde Z_{\mu,2,l}) \hat{D}_{d,k} \sin\left(\frac{2\pi}{\lambda} \Delta d_{k,l}(\mathbf{p}_l)\right) 
\\&+ 2\Im(\tilde Z_{\mu,2,l}) \hat{D}_{d,k} \cos\left(\frac{2\pi}{\lambda} \Delta d_{k,l}(\mathbf{p}_l)\right), \\
\frac{\partial^2 \mathbf{T}_{\mu}(\mathbf{p}_l)}{\partial p_{d,l}^2} &= 
-2\Re(\tilde Z_{\mu,2,l}) \hat{D}_{d,k}^2 \cos\left(\frac{2\pi}{\lambda} \Delta d_{k,l}(\mathbf{p}_l)\right) 
\\&- 2\Im(\tilde Z_{\mu,2,l}) \hat{D}_{d,k}^2 \sin\left(\frac{2\pi}{\lambda} \Delta d_{k,l}(\mathbf{p}_l)\right), \\
\frac{\partial^2 \mathbf{T}_{\mu}(\mathbf{p}_l)}{\partial p_{x,l} \partial p_{y,l}} &= 
-2\Re(\tilde Z_{\mu,2,l}) \hat{D}_{x,k}\hat{D}_{y,k} \cos\left(\frac{2\pi}{\lambda} \Delta d_{k,l}(\mathbf{p}_l)\right) \notag\\
&\quad - 2\Im(\tilde Z_{\mu,2,l}) \hat{D}_{x,k}\hat{D}_{y,k} \sin\left(\frac{2\pi}{\lambda} \Delta d_{k,l}(\mathbf{p}_l)\right).
\end{align}
\end{subequations}

The gradient and Hessian matrix of $\mathbf{T}_{\mu}(\mathbf{p}_l)$ are thus given by
\[
\begin{aligned}
\nabla \mathbf{T}_{\mu}(\mathbf{p}_l) &= 
\begin{bmatrix}
\frac{\partial \mathbf{T}_{\mu}}{\partial p_{x,l}} &
\frac{\partial \mathbf{T}_{\mu}}{\partial p_{y,l}}
\end{bmatrix}^T, \\
\nabla^2 \mathbf{T}_{\mu}(\mathbf{p}_l) &= 
\begin{bmatrix}
\frac{\partial^2 \mathbf{T}_{\mu}}{\partial p_{x,l}^2} & \frac{\partial^2 \mathbf{T}_{\mu}}{\partial p_{x,l} \partial p_{y,l}} \\
\frac{\partial^2 \mathbf{T}_{\mu}}{\partial p_{y,l} \partial p_{x,l}} & \frac{\partial^2 \mathbf{T}_{\mu}}{\partial p_{y,l}^2}
\end{bmatrix}.
\end{aligned}
\]
This completes the derivation.

\section{Derivation of \eqref{56}} \label{appendix:derivation4}
{It can be observed that \( \mathcal{C}_{\text{T},q} \) and \( \mathcal{D}_{\text{T},q} \) share the same structural form, differing only in the beamforming vectors. To streamline the formulation, we introduce a unified notation \( \mathbf{B}_\mu \), defined as
\begin{equation} \small
\mathbf{B}_\mu = |\mathbf{F}_{\text{T},q} \mathbf{w}_\mu|^2 
= Z_{\mu,1} \mathbf{a}_{\text{r}}(\mathbf{p})^H 
\overline{\mathbf{a}}_{\text{T},q}(\mathbf{p}) \mathbf{v}_2 \mathbf{v}_2^H 
\overline{\mathbf{a}}_{\text{T},q}^H(\mathbf{p}) \mathbf{a}_{\text{r}}(\mathbf{p}),
\end{equation}
where \( \overline{\mathbf{a}}_{\text{T},q}(\mathbf{p}) = \operatorname{diag}(\mathbf{a}_{\text{T},q}(\mathbf{p})) \). This expression is structurally identical to \( \mathbf{A}_{\mu,1} \), with only the angular steering vector replaced. Therefore, the derivation is omitted for brevity.}

{By considering a single element position \( \mathbf{p}_l = [p_{x,l}, p_{y,l}]^T \), the partial derivatives of \( \mathbf{B}_\mu(\mathbf{p}_l) \) are given by
\begin{subequations} \small \label{eq:main_equation}
\begin{align}
\frac{\partial \mathbf{B}_\mu(\mathbf{p}_l)}{\partial p_{d,l}} &= -Z_{\mu,1} \hat{D}_{d,q} \sum_{\mathclap{i=1,\; i \neq l}}^{L} \sin \left( \angle s_{i,l,q} + \frac{2\pi}{\lambda} \Delta d_{q,l}(\mathbf{p}_l) \right), \\
\frac{\partial^2 \mathbf{B}_\mu(\mathbf{p}_l)}{\partial p_{d,l}^2} &= -Z_{\mu,1} \hat{D}_{d,q}^2 \sum_{\mathclap{i=1,\; i \neq l}}^{L} \cos \left( \angle s_{i,l,q} + \frac{2\pi}{\lambda} \Delta d_{q,l}(\mathbf{p}_l) \right), \\
\frac{\partial^2 \mathbf{B}_\mu(\mathbf{p}_l)}{\partial p_{x,l} \partial p_{y,l}} &= \frac{\partial^2 \mathbf{B}_\mu(\mathbf{p}_l)}{\partial p_{y,l} \partial p_{x,l}} \notag \\
&= -Z_{\mu,1} \hat{D}_{x,q} \hat{D}_{y,q} \sum_{\mathclap{i=1,\; i \neq l}}^{L} \cos \left( \angle s_{i,l,q} + \frac{2\pi}{\lambda} \Delta d_{q,l}(\mathbf{p}_l) \right),
\end{align}
\end{subequations}
where \( d \in \{x, y\} \). The gradient and Hessian matrix are compactly expressed as
\[
\begin{aligned}
\nabla \mathbf{B}_\mu(\mathbf{p}_l) &= 
\begin{bmatrix}
\frac{\partial \mathbf{B}_\mu}{\partial p_{x,l}} &
\frac{\partial \mathbf{B}_\mu}{\partial p_{y,l}}
\end{bmatrix}^T, \\
\nabla^2 \mathbf{B}_\mu(\mathbf{p}_l) &= 
\begin{bmatrix}
\frac{\partial^2 \mathbf{B}_\mu}{\partial p_{x,l}^2} & \frac{\partial^2 \mathbf{B}_\mu}{\partial p_{x,l} \partial p_{y,l}} \\
\frac{\partial^2 \mathbf{B}_\mu}{\partial p_{y,l} \partial p_{x,l}} & \frac{\partial^2 \mathbf{B}_\mu}{\partial p_{y,l}^2}
\end{bmatrix}.
\end{aligned}
\]}
{
The auxiliary terms are defined as
\[
\begin{aligned}
d_{\text{S},q,l}(\mathbf{p}_l) &= p_{x,l} \sin(\phi_{\text{T},q}) \cos(\psi_{\text{T},q}) + p_{y,l} \sin(\psi_{\text{T},q}), \\
\Delta d_{q,l}(\mathbf{p}_l) &= d_{r,l}(\mathbf{p}_l) - d_{\text{S},q,l}(\mathbf{p}_l), \\
s_{i,l,q} &= \left[ \mathbf{v}_2 \mathbf{v}_2^H \right]_{i,l} \cdot 
e^{-j \frac{2\pi}{\lambda} \left( d_{r,i}(\mathbf{p}_i) - d_{\text{S},q,i}(\mathbf{p}_i) \right)}, \\
\hat{D}_{x,q} &= \frac{2\pi}{\lambda} \left( \sin(\phi_r) \cos(\psi_r) - \sin(\phi_{\text{T},q}) \cos(\psi_{\text{T},q}) \right), \\
\hat{D}_{y,q} &= \frac{2\pi}{\lambda} \left( \sin(\phi_r) - \sin(\phi_{\text{T},q}) \right).
\end{aligned}
\]
}

\bibliographystyle{IEEEtran}
\bibliography{ref}

\end{document}